\documentclass[aps,prb,longbibliography,10pt,twocolumn,superscriptaddress]{revtex4-2}
\usepackage{graphicx} 
\usepackage{amsmath,amsfonts,mathtools,braket}
\usepackage{float}
\usepackage{xcolor}
\usepackage[ruled,vlined,linesnumbered]{algorithm2e} 
\usepackage{soul}
\usepackage{silence}
\def\umphys{
Department of Physics, University of Michigan, Ann Arbor, Michigan 48109, USA
}
\def\umchem{
Department of Chemistry, University of Michigan, Ann Arbor, Michigan 48109, USA
}
\def\ccq{
Center for Computational Quantum Physics, Flatiron Institute, 162 Fifth Avenue, New York, New York 10010, USA
}
\def\upsphys{
Université Paris-Saclay, CNRS, CEA, Institut de physique théorique, 91191, Gif-sur-Yvette, France
}
\def\mstelaviv{
The Raymond and Beverley Sackler Center for Computational Molecular and Materials Science, Tel Aviv University, Tel Aviv 6997801, Israel
}
\def\chemtelaviv{
School of Chemistry, Tel Aviv University, Tel Aviv 6997801, Israel
}

\newcommand{\cc}{c^{*}}
\newcommand{\hc}{\hat{c}}
\newcommand{\hcd}{\hat{c}^{\dagger}}

\begin{document}
\title{Inchworm tensor train hybridization expansion quantum impurity solver}
\author{Yang Yu}
\affiliation{\umphys}
\affiliation{\ccq}
\author{Andr\'e Erpenbeck}
\affiliation{\umphys}
\author{Dominika Zgid}
\affiliation{\umphys}
\affiliation{\umchem}
\author{Guy Cohen}
\affiliation{\mstelaviv}
\affiliation{\chemtelaviv}
\author{Olivier Parcollet}
\affiliation{\ccq}
\affiliation{\upsphys}
\author{Emanuel Gull}
\affiliation{\umphys}
\begin{abstract}
The investigation of quantum impurity models plays a crucial role in condensed matter physics because of their wide-ranging applications, such as embedding theories and transport problems. Traditional methods often fall short in producing accurate results for multi-orbital systems with complex interactions and off-diagonal hybridizations. Recently, tensor-train-based integration and summation techniques have shown promise as effective alternatives. In this study, we use tensor train methods to tackle quantum impurity problems formulated within the imaginary-time inchworm hybridization expansion framework. We identify key challenges in the inchworm expansion itself and its interplay with tensor-train-based methods. We demonstrate the accuracy and versatility of our approach by solving general quantum impurity problems. Our results suggest that tensor-train decomposition schemes offer a viable path toward accurate and efficient multi-orbital impurity solvers.
\end{abstract}
\maketitle

\section{Introduction}

Tensor trains (TT)~\cite{Oseledets_TT_2010,Oseledets_TT_2011}  provide a decomposition of high-order tensors into a sequence of low-order tensors connected through mutual contractions in a chain-like structure.
They were introduced in physics as matrix product states (MPS) in various contexts~\cite{Baxter_MPS_1968,Accardi_MPS_1981,AKLT_MPS_1987,AKLT_MPS_1988,Klumper_MPS_1991,Klumper_MPS_1992,Fannes_MPS_1992,Ostlund_MPS_1995,Dukelsky_MPS_1998} and later applied mathematicians independently developed the same concept under the name of tensor trains~\cite{Oseledets_TT_2009,Oseledets_TT_2010,Oseledets_TT_2011}.
The efficient representation and manipulation of high-dimensional data enabled by TT leads to significant reductions in computational costs and has driven substantial advances in multiple areas of physics. A prominent example is the use of TT to represent wave functions efficiently in low-dimensional quantum lattice systems~\cite{White_DMRG_1992,Schollwock_MPS_2011,Orus_MPS_2014}. TT decomposition has also enabled efficient numerical approaches to differential equations~\cite{Molina_QFT_DE_2022,Haubenwallner_Tensor_HF_2025,Dolgov_FPequations_2012,Ye_Tensor_VPequations_2022,Ye_Tensor_VMequations_2024,Gourianov_Turbulence_2022,Peddinti_Turbulence_2023,Gourianov_Turbulence_2025}, Fourier transforms~\cite{Dolgov_QFT_2012,Molina_QFT_DE_2022,Chen_QFT_2023,Chen_QFT_2024}, function extrapolation~\cite{Matthieu_Extrapolation_2024,Lin_Extrapolation_2023}, and numerical integration~\cite{Khoromskij_QTT_2016,Jolly_Tensor_Orbitals_2024,Finkel_Tensor_Orbitals_2024}.

Recent studies have demonstrated that various fundamental elements in quantum many-body theory, including Green's functions, can be efficiently compressed using the TT decomposition~\cite{Shinaoka_TT_2020,Shinaoka_QTT_2023,Ritter_QTT_2024,Sroda_Tensor_GW_2024,Takahashi_Tensor_GF_2025,Rohshap_Tensor_Parquet_2025}, often in combination with a quantics representation~\cite{Khoromskij_QTT_2011,Khoromskij_QTT_2016}. Moreover, Feynman diagrams have also been shown to admit an efficient representation through the TT decomposition, enabling novel numerical integration schemes for diagrammatic calculation with advantages in both computational speed and accuracy~\cite{Nunez_Learning_2022,Erpenbeck_Tensor_2023,Murray_Tensor_Noneq_2024,Ishida_Tensor_Phonon_2024,Eckstein_Tensor_StrongNoneq_2024,Aaram_Tensor_StrongNoneq_2025,Jeannin_Tensor_Noneq_2025}. For these applications, a specific method for constructing the TT approximation of a high-order tensor, called tensor cross interpolation (TCI), is preferred~\cite{Savostyanov_TCI_2011,Savostyanov_TCI_2014,Dolgov_TCI_2020} to other methods~\cite{White_DMRG_1992,Sakaue_Tensor_Nosiy_2024,Hur_sketch_2023,Aldavero_Chebyshev_2025,Tindall_Tree_2024,Lindsey_Multiscale_2024}. TCI constructs the TT by sampling elements of the high-order tensor rather than requiring knowledge of all tensor elements at once, as in the density matrix renormalization group (DMRG) algorithm~\cite{White_DMRG_1992}.

In this work, we investigate the compatibility of a specific renormalized (bold-line) diagrammatic expansion\textemdash the imaginary-time inchworm hybridization expansion\textemdash with the TT and TCI methods for solving quantum impurity models. Quantum impurity models are the cornerstone of many embedding frameworks designed to solve quantum lattice systems~\cite{Georges_DMFT_1996,Kotliar_DMFT_2006,Zgid_SEET_2017}. A variety of techniques exist for solving quantum impurity models, such as the exact diagonalization~\cite{Caffarel_ED_1994,Capone_ED_2004,Koch_ED_2008,Liebsch_ED_2009,Senechal_ED_2010,Lu_ED_2014,Zgid_CI_2012}, numerical renormalization group~\cite{Weichselbaum_NRG_2007,Bulla_NRG_2008}, and continuous-time quantum Monte Carlo methods~\cite{Gull_CTQMC_2011,Rubtsov_CTINT_2004,Rubtsov_CTINT_2005,Werner_CTHYB_2006,Werner_CTHYB2_2006,Haule_CTHYB_2007,Gull_CTAUX_2008,Gull_CTAUX_2011}. Considerable efforts have been devoted to developing multi-orbital impurity solvers capable of dealing with general interactions and off-diagonal hybridization~\cite{Eitan_inchworm_2020,Li_CDet_2020,Li_Inchworm_2022,Jason_Exponentials_2024,Huang_Exponentials_2025}. One major direction involves the renormalization (boldification) of Feynman diagrams, where the diagrammatic expansion is performed in terms of renormalized (bold) propagators rather than bare propagators. This can lead to a substantial reduction in the required expansion order to achieve a given precision. Bold expansions are often performed using self-consistent iterative methods~\cite{Prokof'ev_Bold_2008,Gull_Bold_2010,Ruegg_XCA_2013,Guy_Bold_2014,Jason_Exponentials_2024,Huang_Exponentials_2025}, which may yield unphysical solutions~\cite{Evgeny_Convergence_2015,Tandetzky_Convergence_2015,Stan_Convergence_2015,Rossi_Convergence_2015,Shafer_Convergence_2016,Gunnarsson_Convergence_2017,Tarantino_Convergence_2017,Vucicevic_Convergence_2018,VanLoon_Convergence_2020,Kim_Convergence_2022,Houcke_Convergence_2024,Eßl_Convergence_2025} or encounter convergence difficulties~\cite{Pokhilko_Convergence_2022,Huang_Exponentials_2025}.
As a reformulation, the inchworm algorithm replaces the iteration procedure with a sequential construction of propagators in a sequence of small time steps.
This method originated in nonequilibrium settings~\cite{Guy_Inchworm_2015,Antipov_Inchworm_2017,Dong_Inchworm_2017,Krivenko_Inchworm_2019,Kleinhenz_Inchworm_2020,Cai_inchworm_2020,Kleinhenz_Inchworm_2022,Erpenbeck_Inchworm_2023,erpenbeck_shaping_2023,Cai_inchworm_2023,Erpenbeck_Inchworm_2024,atanasova_stark_2024} and was later extended to equilibrium problems~\cite{Eitan_inchworm_2020,Li_Inchworm_2022,kim_pseudoparticle_2022,Strand_Inchworm_2024,goldberger_dynamical_2024}. Given the expected reduction in expansion order offered by bold diagrammatic schemes such as the inchworm method, combining them with integration schemes based on TT and TCI methods—which excel at low- to intermediate-dimensional integrals—holds the promise of an efficient and accurate solver.

This paper is organized as follows. In Sec.~\ref{sec:method}, we introduce the methodology, including the inchworm strong-coupling expansion and the fundamentals of TT and TCI. We also discuss key numerical challenges and strategies for integrating TT with the inchworm algorithm. Section~\ref{sec:results} presents our results, providing a systematic analysis of representative models, with particular emphasis on whether the enhanced complexity of resummed diagrams admits a low-rank tensor representation. Finally, Sec.~\ref{sec:conclusion} summarizes our findings and outlines open challenges for the proposed algorithm.

\section{Methods}\label{sec:method}

This work presents a solver for quantum impurity problems, combining the inchworm strong-coupling expansion with TT-based integration and summation techniques. We begin with an overview of the inchworm algorithm, reserving a detailed introduction of the strong-coupling expansion and diagrammatic resummation techniques for Appendix~\ref{appendix:inchworm_derivation}. Next, we introduce the essential concepts of the TT decomposition and the TCI algorithm, referring readers to Refs.~\cite{Dolgov_TCI_2020,Nunez_Learning_2022,Nunez_xfac_2025} for a comprehensive introduction to TCI in this context. Finally, we present the key technical considerations in combining these methodologies.

\subsection{Inchworm algorithm for strong-coupling expansion}\label{subsec:inchworm}
The action of a general fermionic quantum impurity model at inverse temperature $\beta=1/T$ is given by~\cite{Negele_QMBT_2018, Stefanucci_QMBT_2013}
\begin{equation}
    \mathcal{S}_{\mathrm{imp}} = \mathcal{S}_{\mathrm{loc}} + \sum_{12} \cc_{1}\Delta_{12} c_2.
\end{equation}
Here, $\cc$ and $c$ are Grassmann variables corresponding to the fermion creation and annihilation operators $\hcd$ and $\hc$. The composite index $1\equiv(\tau_{1},\iota_{1})$ encodes imaginary time $\tau_{1}$ and additional discrete degrees of freedom $\iota_{1}$ (a composite index of e.g., spin $\sigma_{1}$, orbital/band $l_{1}$, or site indices $r_{1}$). The hybridization function is denoted by $\Delta_{12}\equiv\Delta_{\iota_{1}\iota_{2}}(\tau_{1}-\tau_{2})$, with the first (second) index associated with a creation (annihilation) operator. Summation over $1$, written as $\sum_{1}$, includes integration over continuous variables $\int_{0}^{\beta}\mathrm{d}\tau_{1}$ and  summation over all discrete degrees of freedom $\sum_{\iota_{1}}$. 

A general local action $\mathcal{S}_{\mathrm{loc}}$ takes the form
\begin{equation}
    \mathcal{S}_{\mathrm{loc}} = \sum_{12} \cc_1[\delta(\tau_1-\tau_2)\partial_{\tau_2} + h_{12}]c_2 + \sum_{1234} U_{1423} \cc_{1}\cc_{2} c_{3} c_{4},
\end{equation}
where $h_{12} = h_{\iota_1 \iota_2} \delta(\tau_1-\tau_2)$ is the hopping matrix (which includes a chemical potential term $-\mu \delta_{12}$ in the grand canonical ensemble) and $U_{1423}=U_{\iota_{1}\iota_{4}\iota_{2}\iota_{3}}\delta(\tau_{1}-\tau_{2})\delta(\tau_{1}-\tau_{3})\delta(\tau_{1}-\tau_{4})$ denotes the Coulomb interaction tensor. The corresponding local Hamiltonian reads
\begin{equation}
    \hat{H}_{\mathrm{loc}} = \sum_{\iota_{1}\iota_{2}} h_{\iota_{1}\iota_{2}} \hat{c}^{\dagger}_{\iota_{1}}\hat{c}_{\iota_{2}} + \sum_{\iota_{1}\iota_{2}\iota_{3}\iota_{4}} U_{\iota_{1}\iota_{4}\iota_{2}\iota_{3}} \hat{c}^{\dagger}_{\iota_{1}}\hat{c}^{\dagger}_{\iota_{2}} \hat{c}_{\iota_{3}} \hat{c}_{\iota_{4}}.
\end{equation}

The strong-coupling expansion of the impurity problem is formulated in terms of the renormalized (or ``bold'')  propagator, defined through a bare expansion as [see Appendix~\ref{subsec:strong_coupling_expansion} and Fig.~\ref{fig:hyb}(a) for more details]
\begin{equation}\label{eqn:propagator_main}
\begin{aligned}
\hat{R}(\tau)
&=\sum\limits_{n=0}^{\infty} \sum_{\Phi}\sum\limits_{1\cdots2n} 
\sum_{\mathrm{D}} (-1)^{\sigma} (\Delta)^{n} \\
&\left[\hat{R}^{0}(\tau-\tau_{2n}) \hc_{\iota_{2n}}^{\phi_{2n}} \cdots  \hat{R}^{0}(\tau_{2}-\tau_{1})
    \hc_{\iota_{1}}^{\phi_{1}}\hat{R}^{0}(\tau_{1})
    \right].
\end{aligned}
\end{equation}
Here, $n$ is the expansion order, which corresponds to the number of hybridization functions. The operator type is specified by $\phi_{i}\in \left\{-,+\right\}$, with $\hc^{-}\equiv \hc$ and $\hc^{+}\equiv\hcd$. The set $\Phi=\left\{\phi_1,\cdots,\phi_{2n}\right\}$ runs over all $\binom{2n}{n}$ possible choices of $n$ creation and $n$ annihilation operators. The imaginary time variables are subject to the time-ordering $0<\tau_{1}<\tau_{2}<\cdots<\tau_{2n}<\tau$, and the time integration encoded in $\sum_{1\cdots 2n}$ takes the form    $\int_{0}^{\tau}\mathrm{d}\tau_{1}\int_{\tau_1}^{\tau}\mathrm{d}\tau_{2}\cdots \int_{\tau_{2n-1}}^{\tau}\mathrm{d}\tau_{2n}$. The index $\mathrm{D}$ denotes all distinct assignments of the indices $\left\{1,2,\cdots,2n\right\}$ to the product of $n$ hybridization functions $(\Delta)^{n}$ when $\Phi$ is fixed. The ``bare"  propagator is defined as $\hat{R}^{0}(\tau)\equiv \mathrm{e}^{-\hat{H}_{\mathrm{loc}}\tau}$. The explicit definition of the prefactor $(-1)^{\sigma}$ can be found in Appendix~\ref{appendix:inchworm_derivation}. The impurity partition function is related to the bold propagator through $Z_{\mathrm{imp}}=\operatorname{Tr}[\hat{R}(\beta)]$. By definition, $\hat{R}(0)$ is an identity matrix.

There are several approaches that leverage renormalized propagators to accelerate the convergence of the bare hybridization expansion in Eq.~(\ref{eqn:propagator_main})~\cite{Gull_Bold_2010,gull_numerically_2011,cohen_numerically_2013,cohen_greens_2014,Guy_Bold_2014,haule_dynamical_2010,Guy_Inchworm_2015,Eitan_inchworm_2020}. In the inchworm algorithm, the value of $\hat{R}$ at $\tau$ is computed under the assumption that $\hat{R}$, or frequently an approximation thereof, is already known for all times between $0$ and $\operatorname{max}(\tau_{\mathrm{s}},\tau -\tau_{\mathrm{s}})$, where $\tau_{\mathrm{s}}<\tau$ denotes a ``split" time. The bold propagator at $\tau$ is given by the expression [see Fig.~\ref{fig:tensor_inchworm}(a) and Fig.~\ref{fig:hyb}(b) below for the TT and diagrammatic representations, respectively]
\begin{widetext}
    \begin{equation}\label{eqn:propagator2_main}
    \begin{aligned}
    \hat{R}(\tau)
    &=\sum\limits_{n=0}^{\infty} \sum_{\Phi}\sum\limits_{1\cdots2n} \sum_{\mathrm{D}_{\mathrm{p}}} (-1)^{\sigma} (\Delta)^{n} 
    \left[\hat{R}(\tau-\tau_{2n}) \hc_{\iota_{2n}}^{\phi_{2n}}\cdots \hat{R}(\tau_{q+1}-\tau_{\mathrm{s}}) \hat{R}(\tau_{\mathrm{s}}-\tau_{q}) \cdots \hat{R}(\tau_{2}-\tau_{1})
    \hc_{\iota_{1}}^{\phi_{1}}\hat{R}(\tau_{1})
    \right]\\
    &\equiv\sum\limits_{n=0}^{\infty} \sum_{\Phi}\sum\limits_{1\cdots2n} \sum_{\mathrm{D}_{\mathrm{p}}} \underline{\hat{R}}(\tau|n;\Phi;\iota_{1}\cdots\iota_{2n};\tau_{1}\cdots\tau_{2n};\mathrm{D}_{\mathrm{P}}),
    \end{aligned}
\end{equation}
\end{widetext}
where $0<\tau_{1}<\tau_{2}<\cdots<\tau_{2n}<\tau$ and $\tau_{q}\leq \tau_{\mathrm{s}}\leq \tau_{q+1}$ for certain $q$. Here, the index $\mathrm{D}_{\mathrm{p}}$ runs over a subset of all distinct index assignments of the indices $\left\{1,2,\cdots,2n\right\}$ to the $n$ hybridization functions $(\Delta)^{n}$ when $\Phi$ and all imaginary time variables are fixed. Such valid index assignments $\mathrm{D}_{\mathrm{p}}$ are commonly termed ``inchworm proper" with respect to $\tau_{s}$, for which we provide a precise definition in Appendix~\ref{appendix:inchworm_derivation}. 

In practice, we define a (typically equi-distant) inchworm time grid consisting of $N_{\mathrm{inch}}$ points, $\left\{\tau^{\mathrm{inch}}_{i}| 0=\tau^{\mathrm{inch}}_{0}<\tau^{\mathrm{inch}}_{1} < \cdots < \tau^{\mathrm{inch}}_{N_{\mathrm{inch}}-1} = \beta\right\}$. The calculation proceeds in $N_{\mathrm{inch}}-1$ steps, indexed from $1$ to $N_{\mathrm{inch}}-1$. At the first step ($i=1$), we compute $\hat{R}(\tau_{1}^{\mathrm{inch}})$ using the bare expansion in Eq.~(\ref{eqn:propagator_main}). For subsequent steps ($2\leq i \leq N_{\mathrm{inch}}-1$), we set $\tau_{\mathrm{s}}=\tau^{\mathrm{inch}}_{i-1}$ and $\tau =\tau^{\mathrm{inch}}_{i}$ and perform the bold expansion defined in Eq.~(\ref{eqn:propagator2_main}) with $\hat{R}(\tau\leq\tau^{\mathrm{inch}}_{i-1})$ interpolated, usually with linear or cubic splines, on the grid $\left\{\tau^{\mathrm{inch}}_{0}, \cdots,\tau^{\mathrm{inch}}_{i-1}\right\}$. The expansion of the bold propagator in Eqs.~(\ref{eqn:propagator_main}) and (\ref{eqn:propagator2_main}) is truncated at a maximum expansion order $m$, corresponding to diagrams with up to $m$ hybridization lines.

Using the TT method, we can evaluate the propagator at levels of precision that were previously impractical with Monte Carlo techniques.
We find that the conventional approach of using linear or cubic splines to interpolate the bold propagator $\hat{R}$ is insufficient for achieving high accuracy in some cases. To address this, we introduce a linear-Chebyshev grid to improve the description of the bold propagator. Specifically, in addition to the inchworm grid, for each $i\in \left\{1,\cdots,N_{\mathrm{inch}}-1 \right\}$ we introduce  $n_{\mathrm{Cheby}}+1$ interpolation points $\tau^{\mathrm{Cheby}}_{i,j}$, which satisfy $\tau^{\mathrm{inch}}_{i-1} < \tau^{\mathrm{Cheby}}_{i,1} < \cdots < \tau^{\mathrm{Cheby}}_{i,n_{\mathrm{Cheby}}+1} <\tau^{\mathrm{inch}}_{i}$. These interpolation points $\left\{\tau^{\mathrm{Cheby}}_{i,j}|j=1,\cdots,n_{\mathrm{Cheby}}+1\right\}$ corresponds to the $n_{\mathrm{Cheby}}$-th order Chebyshev nodes between $\tau^{\mathrm{inch}}_{i-1}$ and $\tau^{\mathrm{inch}}_{i}$. We modify the inchworm algorithm such that for each $1\leq i \leq N_{\mathrm{inch}}-1$, we set $\tau_{\mathrm{s}}=\tau^{\mathrm{inch}}_{i-1}$ [not needed if $i=1$] and evaluate $\hat{R}(\tau)$ at $\tau \in \left\{\tau^{\mathrm{Cheby}}_{i,1}, \cdots, \tau^{\mathrm{Cheby}}_{i,n_{\mathrm{Cheby}}+1} ,\tau^{\mathrm{inch}}_{i}\right\}$. The expansion defined in Eq.~(\ref{eqn:propagator2_main}) is performed with $\hat{R}(\tau\leq\tau^{\mathrm{inch}}_{i-1})$ interpolated using Chebyshev polynomials over each subinterval $[\tau^{\mathrm{inch}}_{k-1},\tau^{\mathrm{inch}}_{k}]$ for $k=1,\cdots i-1$. This approach helps yield highly accurate bold propagators, as will be demonstrated in Sec.~\ref{subsec:noninteracting} and Appendix.~\ref{appendix:propagator_noninteracting}.  However, it remains insufficient for more challenging scenarios, such as those discussed in  Sec.~\ref{subsec:Kanamori} and Appendix~\ref{appendix:linear-Cheybshev}.

After obtaining the bold propagator, the one-body Green's function,
\begin{equation}
\begin{aligned}
    G_{j0} &\equiv - \braket{c_{j} \cc_{0}}\\
    &= -\frac{1}{Z_{\mathrm{imp}}} \int \mathcal{D}[\cc,c] \mathrm{e}^{-\mathcal{S}_{\mathrm{imp}}} c_{j} \cc_{0} ,
\end{aligned}
\end{equation}
can be evaluated using the bold propagator~\cite{Guy_Bold_2014} via
\begin{widetext}
\begin{equation}\label{eqn:gf_main}
\begin{aligned}
G_{j0}
& =-\frac{1}{Z_{\mathrm{imp}}} \sum\limits_{n=0}^{\infty} \sum_{\Phi}\sum\limits_{1\cdots2n} 
\sum_{\mathrm{D}_{\mathrm{p}}} (-1)^{\sigma+q} (\Delta)^{n} 
\operatorname{Tr}\left[\hat{R}(\beta-\tau_{2n}) \hc_{\iota_{2n}}^{\phi_{2n}} \cdots \hat{R}(\tau_{q+1}-\tau_{j})\hc_{\iota_{j}} \hat{R}(\tau_{j}-\tau_{q}) \cdots \hat{R}(\tau_{2}-\tau_{1})
    \hc_{\iota_{1}}^{\phi_{1}} \hat{R}(\tau_{1}) \hcd_{\iota_{0}} \right]\\
& \equiv -\frac{1}{Z_{\mathrm{imp}}} \sum\limits_{n=0}^{\infty} \sum_{\Phi}\sum\limits_{1\cdots2n} 
\sum_{\mathrm{D}_{\mathrm{p}}} \underline{G}_{j0}(n;\Phi;\iota_{1}\cdots\iota_{2n};\tau_{1}\cdots\tau_{2n};\mathrm{D}_{\mathrm{P}}).
\end{aligned}
\end{equation}
\end{widetext}
This expression is similar to the inchworm expansion of the bold propagator in Eq.~(\ref{eqn:propagator2_main}). The differences between Eq.~(\ref{eqn:propagator2_main}) and Eq.~(\ref{eqn:gf_main}) are as follows. First, an additional factor $-\frac{1}{Z_{\mathrm{imp}}}$ is added to the expression. Second, the integration over time variables is $\int_{0}^{\beta}\mathrm{d}\tau_{1}\int_{\tau_1}^{\beta}\mathrm{d}\tau_{2}\cdots \int_{\tau_{2n-1}}^{\beta}\mathrm{d}\tau_{2n}$, and $\tau_{q}\leq \tau_{j} \leq \tau_{q+1}$ for certain $q$. Third, the proper diagram is defined with respect to $\tau_{j}$ instead of $\tau_{s}$.  Fourth, the operators $\hc_{\iota_{j}}$ and $\hcd_{\iota_{0}}$ are inserted and a trace is taken. Lastly, there is an additional exponent $q$ in the prefactor $(-1)^{\sigma + q}$, which counts the number of creation or annihilation operators between $\hc_{\iota_{j}}$ and $\hcd_{\iota_{0}}$. In practice, the expansion of one-body Green's function is truncated at a maximum expansion order $m$, corresponding to diagrams with at most $m-1$ hybridization lines.

\subsection{Numerical integration and summation based on tensor trains}\label{subsec:tensor_train}
\begin{figure*}[htb]
\centering
\includegraphics[width=0.9\linewidth]{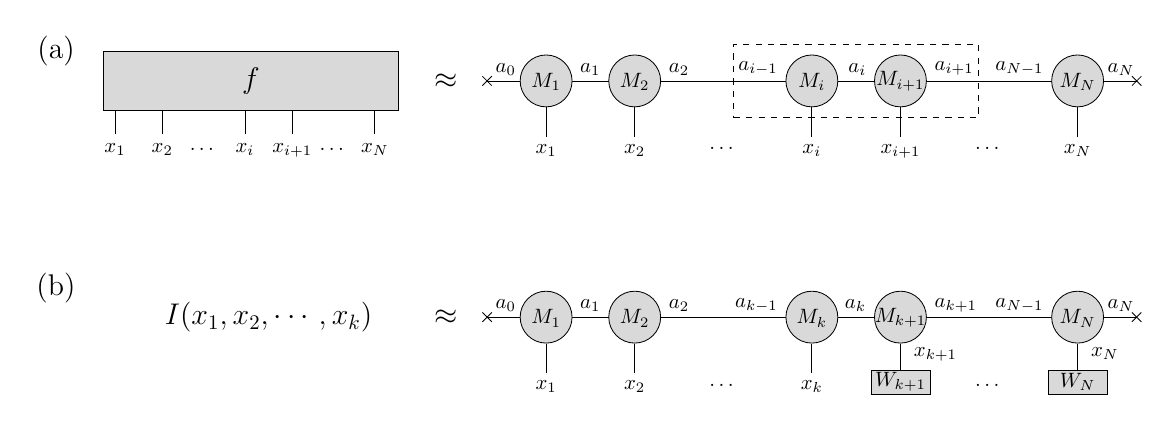}
\caption{Illustration of (a) a tensor train decomposition and (b) numerical integration or summation based on tensor trains. $\left\{x_{i}\right\}$ are the physical indices and $\left\{a_{i}\right\}$ are the virtual indices. Filled circles refer to local three-leg tensors. In (a), the dashed box encloses a local four-leg tensor. In (b), filled squares denote the weights for integration or summation. }
\label{fig:tensor}
\end{figure*}

Within the framework of diagrammatic expansion, the ``curse of dimensionality" inherent to quantum many-body systems manifests itself as the computationally expensive integrations over continuous variables as well as summations over discrete degrees of freedom. Conventional methods rely on quadrature rules and Monte Carlo techniques. Naive quadrature methods are restricted to low-dimensional integrals, while Monte Carlo approaches face rapidly escalating computational costs when higher precision is needed. 

In the last couple of decades, an alternative approach for numerical integration and summation of multivariate functions based on tensor trains~\cite{Oseledets_TT_2010,Oseledets_TT_2011} has emerged in the applied mathematics community~\cite{Savostyanov_TCI_2011,Savostyanov_TCI_2014,Dolgov_TCI_2020}.  This method exhibits remarkable convergence properties, typically achieving relative accuracy $\sim N_{\mathrm{eval}}^{-\alpha} (\alpha \geq 1)$ where $N_{\mathrm{eval}}$ denotes the number of function evaluations, and in some cases demonstrating exponential convergence~\cite{Dolgov_TCI_2020}. Below, we outline the key concepts of this methodology~\footnote{There is an alternative approach for numerical integration that combines TCI with a quantics TT (QTT) representation~\cite{Khoromskij_QTT_2016}. While we do not examine this method in detail in this work, our preliminary analysis indicates reduced efficiency for the present problem: the QTT representation introduces additional physical indices in the TT, typically slowing TCI sampling, while the uniform-grid quadrature used in QTT leads to slower integration convergence. These limitations hinder its effectiveness for higher-order integrals.}; full technical details and implementation aspects can be found in Refs.~\cite{Dolgov_TCI_2020} and ~\cite{Nunez_Learning_2022, Nunez_xfac_2025}.

Consider a multivariate target function $f(x_{1},\cdots,x_{N})$, where some or all variables need to be integrated or summed over. For continuous variables, we first discretize them by selecting appropriate grids. If the continuous variables are to be integrated, the grid points are typically selected according to a one-dimensional quadrature rule with nonuniform weights. This discretization yields a fully discrete order-$N$ tensor of size $d_{1}\times \cdots \times d_{N}$, where each variable $x_{i}$ takes values in a discrete set $\left\{x_{i}^{(1)},\cdots,x_{i}^{(d_{i})}\right\}$. We denote $d\equiv\operatorname{max}_{i}d_{i}$. The TT representation of this multivariate function on the discrete grid takes the form
\begin{equation}\label{eqn:TT}
    f(x_{1},\cdots,x_{N}) = (M_{1})_{a_{0} a_{1}}^{x_{1}} (M_{2})_{ a_{1}a_{2}}^{x_{2}} \cdots (M_{d})_{a_{N-1} a_{N}}^{x_{N}},
\end{equation}
where repeated indices on the right-hand side are implicitly summed over. This decomposition achieves an effective separation of variables through the introduction of tensor contractions between the three-leg (or ``core") tensors $\left\{M_{1},\cdots, M_{d}\right\}$, as shown in Fig.~\ref{fig:tensor}(a). The three-leg tensors possess physical indices $x_i$ that have dimensions $d_{i}$ and virtual indices $a_i$ with dimensions $\chi_{i}$ (referred to as bond dimensions). By definition, $\chi_{0}=\chi_{N}=1$, leaving the corresponding terminal indices uncontracted. The bond dimension $\left\{\chi_{1},\cdots,\chi_{N-1}\right\}$ serves as a control parameter for approximation convergence, with $\chi\equiv \operatorname{max}_{i}\chi_{i}$ defining the rank of the tensor. For many physically relevant functions, a TT representation can be constructed that faithfully captures the original function while exhibiting a rank significantly lower than the theoretical upper bound~\cite{Shinaoka_TT_2020,Shinaoka_QTT_2023,Ritter_QTT_2024,Sroda_Tensor_GW_2024,Takahashi_Tensor_GF_2025,Rohshap_Tensor_Parquet_2025}. Prominent examples are the ground-state wave functions of low-dimensional quantum lattice systems~\cite{Schollwock_MPS_2011,Orus_MPS_2014}.
A key aspect relevant to this work is that Feynman diagrams in many diagrammatic expansions for quantum many-body problems also admit low-rank TT representations~\cite{Nunez_Learning_2022,Erpenbeck_Tensor_2023,Murray_Tensor_Noneq_2024,Ishida_Tensor_Phonon_2024,Eckstein_Tensor_StrongNoneq_2024,Aaram_Tensor_StrongNoneq_2025,Jeannin_Tensor_Noneq_2025}.

While various methods exist for constructing TT representations~\cite{White_DMRG_1992,Sakaue_Tensor_Nosiy_2024,Hur_sketch_2023,Aldavero_Chebyshev_2025,Tindall_Tree_2024,Lindsey_Multiscale_2024}, we employ here the TCI method~\cite{Savostyanov_TCI_2011,Savostyanov_TCI_2014,Dolgov_TCI_2020} for the specific purpose of numerical integration and summation. TCI offers an efficient means to construct the factorized tensor by adaptively sampling only a small subset of the full tensor's elements, termed ``pivots", rather than requiring access to the entire tensor. This adaptive sampling makes TCI particularly suitable for constructing TT representations for functions where evaluation is costly or the full configuration space is intractably large.

The core idea of TCI is to iteratively refine the local four-leg tensors, formed from two neighboring three-leg tensors, by updating corresponding local pivots~\cite{Nunez_xfac_2025}.
This approach is efficient because the algorithm samples only a subspace of the entire tensor at each step. For instance, when updating the four-leg tensor in the dashed box of Fig.~\ref{fig:tensor}(a), the algorithm searches for pivots only in the subspace of $f$, where indices $x_{i}$ and $x_{i+1}$ explore the full $d_{i}\times d_{i+1}$ space, while the remaining indices $(x_{1},\cdots,x_{i-1},x_{i+2},\cdots,x_{N})$ are restricted to a limited subspace determined by previously chosen pivots. The algorithm typically performs sequential sweeps across the TT (from $i=1$ to $i=N-1$, then back from $i=N-1$ to $i=1$), updating the local pivots and local four-leg tensors until a predefined maximum rank $\chi$ or a predefined relative tolerance $\varepsilon$ for the approximation of these local four-leg tensors is achieved. The number of function evaluations of $f$ per sweep scales as $\mathcal{O}(\chi^{2} d^{2} N)$ , while the algebra cost per sweep scales as $\mathcal{O}(\chi^{3} d^{2} N)$~\cite{Nunez_xfac_2025}.

In this work, we utilize the xfac library~\cite{Nunez_xfac_2025}, which implements recent TCI variants. These variants employ the partially rank-revealing LU decomposition for the local updates, offering improved numerical stability compared to the original cross interpolation formulation~\cite{Nunez_xfac_2025}. We note that standard TCI procedures can sometimes encounter ``ergodicity problems", failing to explore parts of the configuration space of the full tensor and thus yielding a misleading result~\cite{Ishida_Tensor_Phonon_2024,Nunez_xfac_2025}. We address this issue by modifying the TCI algorithm to incorporate adaptive random noise, the details of which are provided in Appendix~\ref{appendix:ergodicity}.

Once the TT approximation has been obtained via TCI, the subsequent high-dimensional integration or summation problem can be reformulated as a product of one-dimensional integrations or summations, which can be performed efficiently. Suppose we have ordered the function $f(x_{1},x_{2},\cdots,x_{N})$  such that the first $k$ indices are retained, while the remaining $N-k$ indices need to be integrated or summed over. For continuous variables $x_{i}$ that needed to be integrated over, we discretize them on a quadrature grid with corresponding quadrature weights $W_{x_{i}}$\footnote{While the grid used for constructing the TT approximation can, in principle, differ from the grid on which the approximation is subsequently evaluated, we use the same grid for both construction and evaluation here.}. For, the discrete variables $x_{j}$ that need to be summed over, we assign uniform weights $W_{x_{j}}=\left\{1,\cdots,1\right\}$. The combined summation and integration can then be approximated as
\begin{equation}\label{eqn:TT_sum}
\begin{aligned}
I(x_{1},\cdots,x_{k}) 
&\equiv \sum_{(k+1)\cdots N}  f(x_{1},\cdots,x_{N})\\
&\approx (M_{1})_{a_{0} a_{1}}^{x_{1}} \cdots 
    (M_{k})_{a_{k-1} a_{k}}^{x_{k}}\\
    & \cdot[(M_{k+1})_{a_{k} a_{k+1}}^{x_{k+1}} W_{x_{k+1}}]\cdots [(M_{d})_{a_{N-1} a_{N}}^{x_{N}} W_{x_{N}}].
\end{aligned}
\end{equation}
This operation is illustrated graphically in Fig.~\ref{fig:tensor}(b), where the square blocks denote the weights for integration or summation, and the internal indices can be contracted sequentially from right to left.

\subsection{Application of tensor trains to the inchworm hybridization-expansion}\label{subsec:TT+Inchworm}
\begin{figure*}[htb]
\centering
\includegraphics[width=0.8\linewidth]{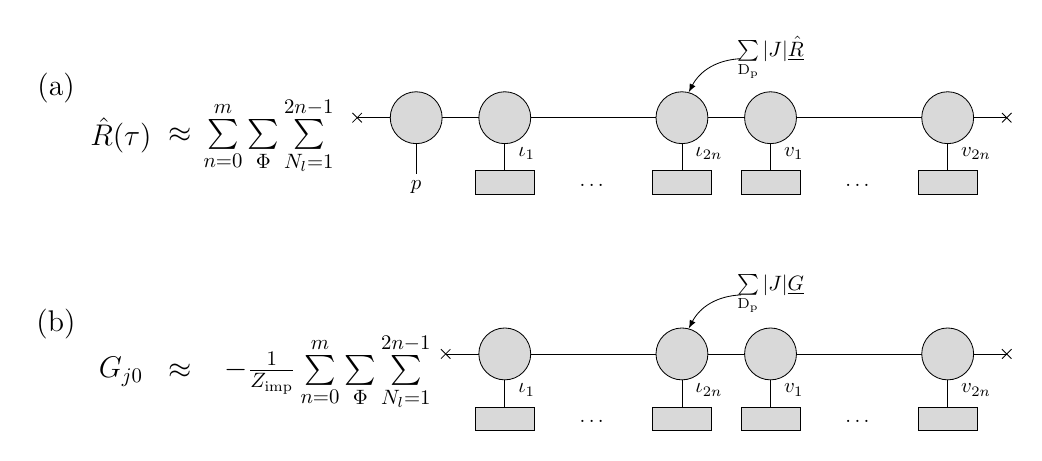}
\caption{Illustration of the TT configurations used in this work for (a) the bold propagator $\hat{R}(\tau)$ and (b) the Green's function $G_{j0}$. The indices $\left\{\iota_{i}\right\}$ encode composite discrete degrees of freedom (e.g., spin, orbital, or site), while $\left\{v_{i}\right\}$ correspond to transformed variables obtained by mapping time-ordered imaginary time variables $\left\{\tau_{i}\right\}$  from a simplex to a hypercube. In (a), the index $p$ belongs to a one-dimensional array mapped from the two matrix indices of $\hat{R}(\tau)$.}
\label{fig:tensor_inchworm}
\end{figure*}

We now revisit the expressions in Eq.~(\ref{eqn:propagator2_main}) for the bold propagator (with analogous considerations applying to Eq.~(\ref{eqn:propagator_main})) and Eq.~(\ref{eqn:gf_main}) for the Green's function. To efficiently evaluate the integrations and summations in these expressions using a TT representation, several key questions must be addressed.

\subsubsection{Different approaches to performing integration or summation}

Eqs.~(\ref{eqn:propagator2_main}) and (\ref{eqn:gf_main}) require the integration or summation over the expansion order $n$, the operator configurations $\Phi$, the degrees of freedom of the model $\left\{1,\cdots, 2n\right\}$, and the diagram topology $\mathrm{D}_{\mathrm{p}}$.
For each integration or summation operation, three distinct approaches are possible. The first approach incorporates the integration or summation into the target function represented by the TT, meaning that the integration or summation is performed during each sampling process of the TCI algorithm. While this treatment may modify the TT rank in an {\it a priori} unknown manner, it invariably slows down the construction of TT, raising the number of function calls per TCI sweep from $\mathcal{O}(\chi^2 d^2 N)$ to $\mathcal{O}(\chi^2 d^3 N)$, assuming a summation of size $d$ is included in the target function. The second approach handles the integration or summation at the TCI level as shown in Eq. (\ref{eqn:TT_sum}), where the introduction of additional physical indices typically increases the TT rank due to the enlarged configuration space of the TT. The third alternative excludes the integration or summation operations from the TT framework entirely, performing them explicitly after completion of the TT operations. 

For the index $\mathrm{D}_{\mathrm{p}}$, we explicitly choose to include them as part of the target function for the TT. The reason is that for different combinations of $(\Phi,1\cdots2n)$, the number of valid diagram topologies varies. Thus, if we were to treat $\mathrm{D}_{\mathrm{p}}$ as a physical index of the TT with a fixed size, we would need to enlarge the configuration space to accommodate all possibilities, assigning zero weight to invalid diagram topologies. This would result in a physical index of excessively large dimension, leading to a potentially high-rank TT in order to maintain accuracy. Moreover, as discussed in Appendix~\ref{appendix:ergodicity}, naively introducing many invalid configurations as zeros or constants makes it easy to trigger ergodicity problems. For these reasons, we incorporate $\mathrm{D}_{\mathrm{p}}$ into the target function of the TT.

The integration and summation over the index set $\left\{1,\cdots, 2n\right\}$ constitute the most computationally expensive part of the calculation. Therefore, these indices are explicitly incorporated as separate physical indices in the TT representation. It is worth noting that the TT-based method provides a distinct advantage over some other recently developed bold expansion algorithms~\cite{Strand_Inchworm_2024,Jason_Exponentials_2024,Huang_Exponentials_2025}, which primarily handle continuous-variable integration efficiently and leave the summation over discrete variables intact. By effectively compressing the information encoded in the discrete indices $\left\{\iota_{1},\cdots, \iota_{2n}\right\}$, the TT-based approach can alleviate the expensive summation over these discrete indices.

We treat the indices $\Phi$ through explicit summation rather than incorporating them into the TT representation. This choice stems from two key considerations for the $\Phi$ index: First, naively encoding $\Phi$ in the TT would require expanding the configuration space to $2^{2n}$ states (by assigning each $\phi_i$ as a binary physical index), significantly larger than the physical $\binom{2n}{n}$ configurations. Also, as discussed in Appendix~\ref{appendix:ergodicity}, introducing additional nonphysical configurations as zeros or constants could lead to severe ergodicity problems. Second, while an alternative encoding using a single index of dimension $\binom{2n}{n}$ is possible, this approach typically requires much higher TT ranks for convergence and substantially slows down local update operations.

For the diagram order $n$, we also use explicit summation because varying $n$ alters both the number of $\iota$ indices and $\tau$ indices, thereby changing the fundamental structure of the TT representation itself.

\subsubsection{Special treatment for time integration}

The index set $\left\{1,\cdots, 2n\right\}$ requires separate treatment of its two distinct components. The discrete indices $\left\{\iota_{1},\cdots,\iota_{2n}\right\}$ can be straightforwardly included as physical indices in the TT. The time indices $\left\{\tau_{1}, \cdots,\tau_{2n} \right\}$ present a challenge for TT decomposition. When these indices are not time-ordered, the resulting discontinuities that occur whenever any $\tau_{i}$ crosses another severely degrade the compressibility of the target function by TT~\cite{Nunez_Learning_2022}. To overcome this difficulty, we employ a time-ordered representation of the expansion, as specified in Eqs.~(\ref{eqn:propagator2_main}) and (\ref{eqn:gf_main}). An additional complication arises at $\tau_{\mathrm{mid}}$ [where $\tau_{\mathrm{mid}}=\tau_{\mathrm{s}}$ for Eq.~(\ref{eqn:propagator2_main}) and $\tau_{\mathrm{mid}}=\tau_{j}$ for Eq.~(\ref{eqn:gf_main})] because any $\tau_{i}$ crossing this point creates discontinuities that reduce compressibility, as discussed in Refs~\cite{Erpenbeck_Tensor_2023} and~\cite{Matsuura_Tensor_Weak_2025}. We therefore reformulate the original time-ordered integral 
\begin{equation}
\int_{\tau_{\mathrm{min}}}^{\tau_{\mathrm{max}}}\mathrm{d}\tau_{1}\int_{\tau_1}^{\tau_{\mathrm{max}}}\mathrm{d}\tau_{2}\cdots \int_{\tau_{2n-1}}^{\tau_{\mathrm{max}}}\mathrm{d}\tau_{2n} \equiv \int \mathrm{d} \vec{\tau}
\end{equation}
[with $\tau_{\mathrm{min}}=0$ and $\tau_{\mathrm{max}}=\tau$
for Eq.~(\ref{eqn:propagator2_main}) and $\tau_{\mathrm{max}}=\beta$ for Eq.~(\ref{eqn:gf_main})] into a piecewise-ordered form:
\begin{equation}
    \sum_{N_{l}=0}^{2n} \int_{\tau_{\mathrm{min}}}^{\tau_{\mathrm{mid}}}\mathrm{d}\tau_{1}\cdots \int_{\tau_{N_{l}-1}}^{\tau_{\mathrm{mid}}}\mathrm{d}\tau_{N_{l}}  \int_{\tau_{\mathrm{mid}}}^{\tau_{\mathrm{max}}}\mathrm{d}\tau_{N_{l}+1}  \cdots \int_{\tau_{2n-1}}^{\tau_{\mathrm{max}}}\mathrm{d}\tau_{2n},
\end{equation}
where $\tau_{\mathrm{min}}<\tau_{1}<\cdots < \tau_{N_{l}}< \tau_{\mathrm{mid}}$ and $\tau_{\mathrm{mid}}< \tau_{N_{l}+1}<\cdots  < \tau_{2n}< \tau_{\mathrm{max}}$. The requirement for proper diagrams further constrains the number of time indices preceding $\tau_{\mathrm{mid}}$ to the range $1\leq N_{l}\leq 2n-1$. 

A TT representation inherently necessitates mutually independent physical indices. However, this condition is incompatible with the imaginary time indices $\left\{\tau_{1},\cdots,\tau_{2n}\right\}$ in Eqs.~(\ref{eqn:propagator2_main}) and (\ref{eqn:gf_main}), which are explicitly constrained by temporal ordering. To resolve this incompatibility, we implement a variable transformation following Ref.~\cite{Erpenbeck_Tensor_2023}, mapping the original time indices $\tau_{1},\cdots,\tau_{2n}$ to a new set of independent variables $v_{1},\cdots,v_{2n} \in [0,1]$~\footnote{Ref.~\cite{Erpenbeck_Tensor_2023} suggested one may find potential improvements to this mapping scheme. Indeed, alternative approaches such as the simplex-to-hypercube transformation used in Ref.~\cite{Strand_Inchworm_2024} offer more uniform distributions. However, such mappings introduce some computational overhead, requiring both the evaluation of high-order roots during sampling and modified integration schemes to handle the resulting singularities in the integrand. For the diagram orders considered in this work, we find that the choice of mapping is not the dominant bottleneck.}:
\begin{equation}
    \begin{aligned}
            v_{1} &= \frac{\tau_{1}-\tau_\mathrm{min}}{\tau_{\mathrm{mid}}-\tau_\mathrm{min}}\\ 
            v_{i} &= \frac{\tau_{i} - \tau_{i-1}}{\tau_\mathrm{mid}-\tau_{i-1}} \qquad 2 \leq i \leq N_{l} \\
            v_{N_{l}+1} &= \frac{\tau_{N_{l}+1}-\tau_\mathrm{mid}}{\tau_{\mathrm{max}}-\tau_\mathrm{mid}}\\ 
            v_{k} &= \frac{\tau_{k} - \tau_{k-1}}{\tau_\mathrm{max}-\tau_{k-1}} \qquad N_{l}+2 \leq k \leq 2n
    \end{aligned}
\end{equation}
where the corresponding Jacobian is 
\begin{equation}
\begin{aligned}
    J &=(\tau_\mathrm{mid}-\tau_\mathrm{min})(\tau_\mathrm{mid} - \tau_{1})\cdots (\tau_\mathrm{mid}-\tau_{N_{l}-1})\\
    & \times (\tau_\mathrm{max}-\tau_\mathrm{mid})(\tau_\mathrm{max} - \tau_{N_{l}+1})\cdots (\tau_\mathrm{max}-\tau_{2n-1}).
\end{aligned}
\end{equation}
The time-ordered integration over imaginary time indices can then be rewritten as 
\begin{equation} 
\int \mathrm{d} \vec{\tau} = \sum_{N_{l}=1}^{2n-1} \int_{0}^{1}\mathrm{d}v_{1}\cdots \int_{0}^{1}\mathrm{d}v_{2n} |J|.
\end{equation}
The transformed variables $\left\{v_1,\cdots,v_{2n}\right\}$ admit straightforward discretization through standard quadrature grid, enabling their direct incorporation into the TT framework as described in Sec.~\ref{subsec:tensor_train}.  In this work, we employ the Gauss-Kronrod quadrature rule of order $N_{\mathrm{GK}}$, which corresponds to $N_{\mathrm{GK}}$ Kronrod points for numerical integration.

\subsubsection{Indices that require no integration or summation}

Besides the indices to be integrated or summed over, the TT representation allows for additional flexibility in handling various types of indices that do not need to be integrated or summed over, as discussed in Sec.~\ref{subsec:tensor_train}. For Eq.~(\ref{eqn:propagator2_main}), potential candidates for such physical indices include the matrix elements indices of $\underline{\hat{R}}$ and the $\tau$ index enumerating time grid points. Our numerical analysis reveals that incorporating the matrix index of $\underline{\hat{R}}$, denoted as $p$ (a one-dimensional array mapped from the two matrix indices of $\hat{R}$), as a single physical index leads to a marginal increase in the required TT rank for fixed precision. Including the $\tau$ index in the TT sometimes increases the necessary rank of the TT, depending on the specific shape of the propagator.  This effect can hinder achieving precise results for the propagator. Consequently, we adopt a conservative approach where only the matrix element index $p$ is treated as a physical index while maintaining a separate TT for each $\tau$ value.

A similar consideration applies to $\underline{G}$ in Eq.~(\ref{eqn:gf_main}), which carries indices $\tau_{j}$ and $\iota_{j}$.  While high-temperature regimes permit their inclusion as physical indices in TT without significant complications, we observed that this treatment demands substantially larger ranks for accurate results at low temperatures. This temperature-dependent behavior likely stems from the increasingly disparate values of the Green's function across different $\tau_{j}$ and $\iota_{j}$ at lower temperatures. We therefore exclude both indices from the TT.

\subsubsection{The ordering of the physical indices}

Having established the set of physical indices for the TT representation, we now address the question of index ordering. While previous studies have investigated the impact of index ordering on TT performance~\cite{Nunez_Learning_2022}, our analysis reveals that for the present problem, the specific ordering has minimal effect on the resulting rank of the TT. Through explicit verification using the DMRG algorithm for low-order diagrams, we observe comparable performance across different orderings. Nevertheless, the particular ordering $(\iota_{1},\cdots,\iota_{2n},v_{1},\cdots,v_{2n})$ seems to be marginally more efficient than alternative configurations and is thus used in this work.

\subsubsection{Summary}

In summary, when combining the TT with the inchworm algorithm, the specific scheme employed in this work follows Fig.~\ref{fig:tensor_inchworm}(a) for the propagator $\hat{R}$ at each time $\tau$ and follows Fig.~\ref{fig:tensor_inchworm}(b) for the Green's function $G$ at each $\tau_{j}$ and $\iota_{j}$. The explicit summation on the right-hand side is parallelized in a straightforward manner. In the initial step of the inchworm algorithm, Eq.~(\ref{eqn:propagator_main}) is employed rather than Eq.~(\ref{eqn:propagator2_main}). The TT construction follows a similar structure to Fig.~\ref{fig:tensor_inchworm}(a) but avoids the complexity from the piecewise-ordered time integration.

\section{Results}\label{sec:results}
In this section, we discuss the successes and challenges of the algorithm proposed in Sec.~\ref{subsec:TT+Inchworm}, which is based on the inchworm resummation scheme and the TT technique. In Sec.~\ref{subsec:noninteracting}, we benchmark the algorithm against a spinless model, demonstrating excellent agreement with exact results. In Sec.~\ref{subsec:Kanamori}, we illustrate the algorithm's capability to handle multi-orbital impurity models with general interactions and off-diagonal hybridization. In Sec.~\ref{subsec:convergence}, we investigate the convergence behavior of the algorithm at different temperatures, different interaction strengths, and for different bath configurations. 

The major numerical parameters to be specified for the algorithm include: the number of inchworm grid points $N_{\mathrm{inch}}$, the order of the Chebyshev polynomial for interpolation used within each linear inchworm grid interval $n_{\mathrm{Cheby}}$, the number of Kronrod points for numerical integration $N_{\mathrm{GK}}$, the rank of TT $\chi$, the relative tolerance of TCI $\varepsilon$, and the maximum expansion order $m$. Throughout this work, we set $n_{\mathrm{Cheby}}=9$, $N_{\mathrm{GK}}=15$ and $\varepsilon=10^{-14}$. The remaining numerical parameters, together with all relevant physical parameters, will be specified in the sections for each example.

All calculations in this work are deterministic. Variations in symbol sizes and line thicknesses in all subsequent figures are employed solely for visual distinction and do not convey uncertainty.

    \subsection{Noninteracting limit}\label{subsec:noninteracting}
    
    \begin{figure}[tb]
        \centering
        \includegraphics[width=1\linewidth]{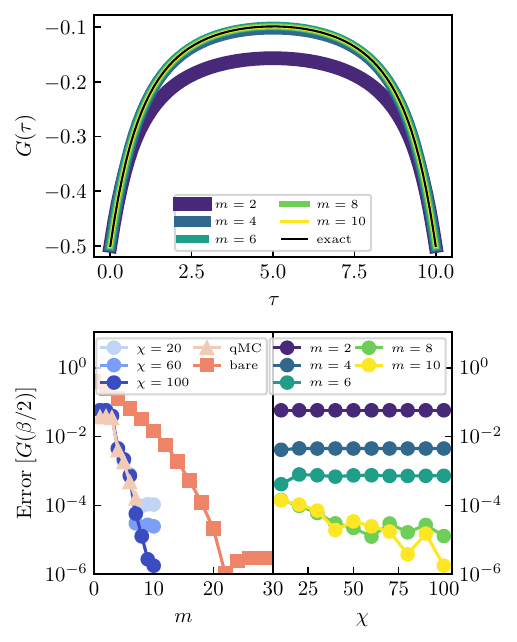}
        \caption{
        Green's functions of the spinless non-interacting model. Top panel: Green's functions for different maximum expansion orders $m$ (color-coded) compared to the exact Green's function (black line). Bottom panel: Absolute error of the Green's function at $\tau=\beta/2$ as a function of (left) maximum expansion order $m$ (circles, color-coded by $\chi$) and (right) rank $\chi$ (color-coded by $m$).
        Reference data include quasi-Monte Carlo inchworm expansion results (triangles) from Ref.~\cite{Strand_Inchworm_2024} and TT-based bare expansion results (squares) from Ref.~\cite{Erpenbeck_Tensor_2023}.
        }
        \label{fig:spinless_Bethe_G}
    \end{figure}
    
    We begin by analyzing the algorithm for one of the simplest impurity models—a noninteracting spinless system described by the action
    \begin{equation}
        \mathcal{S}_{\mathrm{imp}}= \int_0^\beta \mathrm{d} \tau \mathrm{d} \tau^{\prime} \cc(\tau)\left[\delta\left(\tau-\tau^{\prime}\right) \partial_\tau + \Delta\left(\tau-\tau^{\prime}\right)\right] c\left(\tau^{\prime}\right).
    \end{equation}
    We choose a hybridization function
    \begin{equation}\label{eqn:hyb}
        \Delta(\tau)= - \int \mathrm{d}\omega \frac{\Gamma(\omega)\mathrm{e}^{-\tau\omega}}{1+\mathrm{e}^{-\beta\omega}}
    \end{equation}
    corresponding to a continuous bath with semicircular-shape spectrum $\Gamma(\omega) = \sqrt{4t^2-\omega^2}/(2 \pi t^2)$, where $ -2t \leq \omega \leq 2t$ and $t=1$ sets the energy scale throughout the rest of the text. For this specific non-interacting impurity model, the exact Green’s function and the partition function are known. This makes the model an ideal, nontrivial benchmark for strong-coupling expansion-based impurity solvers~\cite{Werner_CTHYB_2006}. 

    Figure~\ref{fig:spinless_Bethe_G} displays the Green's function for the spinless model at $\beta=10$,  computed with the propagator evaluated with $N_{\mathrm{inch}}=11$, $\chi=100$, and $m=7$. The corresponding analysis for the propagator is detailed in Appendix~\ref{appendix:propagator_noninteracting}.  The top panel of Fig.~\ref{fig:spinless_Bethe_G} compares $G(\tau)$  at various maximum expansion orders $m$ (solid lines, evaluated with $\chi=100$) with the exact solution (solid black line). Rapid convergence is observed here despite the non-interacting nature of the model, which typically poses challenges for strong-coupling expansions. The bottom panel presents the absolute error at $\tau=\beta/2$ versus the maximum expansion order $m$ (left) and the rank of TT $\chi$ (right). For reference, we include the published results: quasi-Monte Carlo inchworm expansion data from Ref.~\cite{Strand_Inchworm_2024} (triangles) and TT bare expansion data with $\chi=120$ from Ref.~\cite{Erpenbeck_Tensor_2023} (squares) in the lower-left panel. Our method achieves comparable or superior precision to these approaches.
    In the lower-right panel, we observe that for $m\leq 6$, convergence requires only modest rank $\chi\sim20$. 
    This is consistent with recent publications demonstrating the suitability of TT-based methods for low-order resummation techniques \cite{Eckstein_Tensor_StrongNoneq_2024, Aaram_Tensor_StrongNoneq_2025}.
    For higher expansion orders, however, the TT rank required to maintain accuracy grows rapidly, highlighting the necessity for methodological advancements to enable feasible higher-order tensor train calculations.
    
    \subsection{Multi-orbital impurity problems}\label{subsec:Kanamori}
    \begin{figure}[tb]
        \centering
        \includegraphics[width=0.95\linewidth]{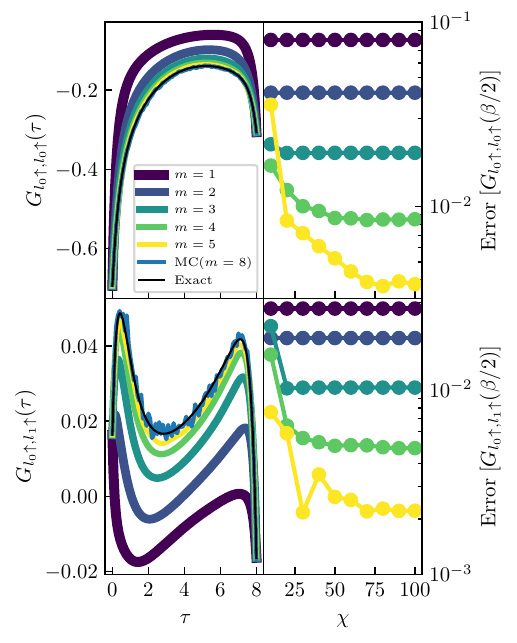}
        \caption{Green's functions of the two-orbital spinful model. Top panels: Diagonal component. Bottom panels: Off-diagonal component. Left panels: Green's functions calculated with rank $\chi=100$. Results for different expansion orders $m$ are shown in color, compared to the exact solution (black line) and Monte Carlo data ($m=8$) from Ref.~\cite{Eitan_inchworm_2020}. Right panels: Absolute error at $\tau=\beta/2$ as a function of rank $\chi$, using the same color scheme as the left panels. }
        \label{fig:Kanamori}
    \end{figure}

    After demonstrating the ability of the TT method to obtain highly precise results for the simple non-interacting case, we now consider a two-orbital spinful interacting impurity model with a local Hamiltonian of Kanamori form~\cite{Kanamori_1963}, relevant for Hund's metal physics~\cite{Werner_SpinFreezing_2008,Georges_Hund_2013}:
    \begin{equation}
    \begin{aligned}
    \hat{H}_{\mathrm {loc }} 
    &= U \sum_{l\in\{l_{0},l_{1}\}} \hat{n}_{l \uparrow} \hat{n}_{l \downarrow}\\
    & +\sum_{\sigma, \sigma^{\prime} \in\{\uparrow, \downarrow\}}\left(U^{\prime}-J_H \delta_{\sigma \sigma^{\prime}}\right) \hat{n}_{l_{0} \sigma} \hat{n}_{l_{1} \sigma^{\prime}} \\
    & +J_H\left(\hcd_{l_{0} \uparrow} \hcd_{l_{0} \downarrow} \hc_{l_{1} \downarrow} \hc_{l_{1} \uparrow}+\hcd_{l_{0} \uparrow} \hcd_{l_{1} \downarrow} \hc_{l_{0} \downarrow} \hc_{l_{1} \uparrow}+h.c.\right),
    \end{aligned}
    \end{equation}
    where $\hat{n}_{l_{i}\sigma_{i}}\equiv\hat{c}_{l_{i}\sigma_{i}}^{\dagger}\hat{c}_{l_{i}\sigma_{i}}$ denotes the density operator. The first row represents intra-orbital Coulomb repulsion, where we set the interaction strength to be $U=2$. The second row describes the inter-orbital interaction for anti-parallel ($\sigma \neq \sigma'$) and parallel ($\sigma=\sigma'$) spin, where $U'=U-2J_H$ and we set $J_{H}=0.2$. The last row represents the ``pairing hopping" and ``spin exchange" processes. The chemical potential is set to $\mu = (3U-5J_{H}-3)/2=1$ and the inverse temperature is set to $\beta=8$. The hybridization function is set to $\Delta_{l_{i}\sigma_{i},l_{j}\sigma_{j}}(\tau)= \delta_{\sigma_{i}\sigma_{j}}\sum_{k} (1+0.25 \delta_{l_{i}l_{j}})[\Delta_{e_{0}}(\tau)+\Delta_{e_{1}}(\tau)]$ with $\Gamma_{e_{k}}(\omega)=\delta(\omega-e_{k})$, $e_{0}=-2.3$ and $e_{1}=2.3$ [The relation between $\Gamma_{e_{k}}(\omega)$ and $\Delta_{e_{k}}(\tau)$ follows Eq.~(\ref{eqn:hyb})]. Such off-diagonal hybridization typically leads to a severe sign problem in the Monte Carlo-based bare expansion~\cite{Gull_CTQMC_2011,Werner_CTHYB_2006,Werner_CTHYB2_2006,Haule_CTHYB_2007,Eitan_inchworm_2020}.

    For this specific model, unlike the spinless model discussed in Sec.~\ref{subsec:noninteracting}, the use of a linear-Chebyshev grid still requires a large number of linear grid points to accurately represent the bold propagator (see Appendix~\ref{appendix:linear-Cheybshev}). This leads to significant computational cost and potentially large error propagation in the inchworm scheme, both of which hinder the accurate determination of the Green's function. To circumvent these issues in our analysis for the Green's function, we employ the exact bold propagator obtained from exact diagonalization on a dense grid ($N_{\mathrm{inch}}= 501$) and compute the Green's function accordingly.

    Figure~\ref{fig:Kanamori} illustrates the convergence behavior of the Green's function as a function of the maximum expansion order $m$ for the diagonal (top panels) and off-diagonal (bottom panels) components. For comparison, we also display the Monte Carlo inchworm results from Ref.~\cite{Eitan_inchworm_2020} for the same model, truncated at expansion order $m=8$. As shown in the right panels of Fig.~\ref{fig:Kanamori}, a modest rank $\chi=20$ suffices to converge at orders $m\leq 3$. However, the required rank grows rapidly for higher orders ($m=4$ and $m=5$). In contrast to the spinless case in Sec.~\ref{subsec:noninteracting}, the configuration space to be compressed via the TT decomposition is substantially larger due to the additional choices of the spin-orbital indices, which naturally necessitates a higher rank and renders computations at even higher expansion orders computationally prohibitive. In our current unoptimized implementation of the algorithm, computing a single Green's function data point for $m=5$ and $\chi=100$ requires approximately 500 core-hours.

    \subsection{Convergence analysis}\label{subsec:convergence}
    \begin{figure}[tb]
        \centering
        \includegraphics[width=1\linewidth]{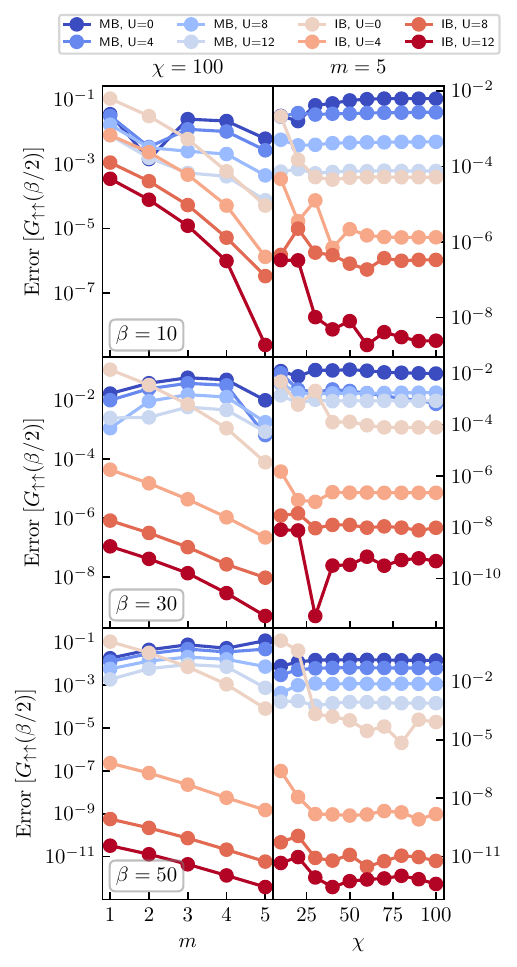}
        \caption{Absolute error in the Green's functions of the single-orbital spinful model at $\tau=\beta/2$ for different inverse temperatures. Results are shown for $\beta=10$ (top), $\beta=30$ (middle) and $\beta=50$ (bottom). Left panels: Error as a function of maximum expansion order $m$ at fixed rank $\chi=100$. Right panels: Error as a function of rank $\chi$ at fixed expansion order $m=5$. Different interaction strengths $U$ and bath configurations are distinguished by color.}
        \label{fig:convergence}
    \end{figure}
    After demonstrating the performance of our methodology on specific representative models, we now provide a comprehensive overview of the TT-based algorithm's performance across a wide range of parameter regimes, including various temperatures, interaction strengths, and bath configurations.
    We focus on a single-orbital spinful interacting model with the local Hamiltonian given by
    \begin{equation}
        \hat{H}_{\mathrm{loc}} = U \hat{n}_{\uparrow} \hat{n}_{\downarrow}.
    \end{equation}
     The hybridization function takes the form
     $\Delta_{\sigma_{i}\sigma_{j}}(\tau)= \delta_{\sigma_{i}\sigma_{j}}\sum_{k} V^{*}_{e_{k}}V_{e_{k}}\Delta_{e_{k}}(\tau)$ where $\Gamma_{e_{k}}(\omega)=\delta(\omega-e_{k})$ (with $\Delta_{e_{k}}(\tau)$ related to $\Gamma_{e_{k}}(\omega)$ through Eq.~(\ref{eqn:hyb})). We examine two distinct bath configurations: an insulating bath (IB) characterized by discrete states at energies $e_{k}\in\left\{-1.0,1.0\right\}$ with corresponding coupling $V_{e_{k}}\in\left\{0.6,0.6\right\}$, and a metallic bath (MB) with energies at $e_{k}\in\left\{-1.0,0.0,1.0\right\}$ with corresponding coupling $V_{e_{k}}\in\left\{0.6,0.5,0.6\right\}$. The chemical potential is set to $\mu=U/2$ to maintain half-filling conditions. We employ the exact bold propagator (with $N_{\mathrm{inch}}=\beta+1$) for the calculation of the Green's function to avoid amplifying errors from inchworm propagation.

     Figure~\ref{fig:convergence} shows the absolute error of the Green’s function at $\tau = \beta/2$, which we use as a metric to assess the precision of our results\textemdash consistent with our analysis of the spinless non-interacting model\textemdash across different inverse temperatures: 
     $\beta=10$ (top), $\beta=30$ (middle) and $\beta=50$ (bottom). The left panels show the error as a function of maximum expansion order $m$ with fixed rank $\chi=100$, while the right panels display the error as a function of rank $\chi$ at fixed expansion order $m=5$. Results are shown for different interaction strengths $U$ and bath configurations, distinguished by color.

     The left panels reveal that systems with insulating bath configurations (red curves) converge significantly faster with expansion order than their metallic bath counterparts (blue curves). In the meantime, increasing the Hubbard interaction $U$ improves convergence, which is consistent with expectations from hybridization expansion frameworks. Notably, at $\beta=50$ (bottom left panel) with weak interactions, the expansion does not show a trend that indicates convergence at $m=5$, a behavior we have also verified through Monte Carlo calculations. This detrimental effect of metallic bath configurations on the convergence properties of bold hybridization expansions\textemdash likely caused by zero-energy bath fluctuations enabling many concurrent hybridization events\textemdash has also been recently noted in Ref.~\cite{Huang_Exponentials_2025}.
     
     The right panels show that the required TT rank for convergence shows relatively weak dependence on temperature $\beta$ and interaction strength $U$. While the metallic bath exhibits better convergence compared to the insulating bath, the overall accuracy remains primarily limited by the order truncation.
     Still, we note that the rank required for an accurate TT representation depends not only on the diagrammatic structure but also on the choice of input parameters in the calculation, such as the hybridization function. This consideration could become important when extending TT-based methods to parameter regimes without benchmarks.
     For all configurations considered here, convergence is typically achieved with rank $\chi\leq50$.

    Based on the observations presented above and in Sec.~\ref{subsec:Kanamori}, we conclude that a key challenge for generic strong-coupling impurity solvers arises from the shift of the order distribution toward higher expansion orders. This shift may result from several factors, including model complexity, lower temperatures, and—most critically—a bath hybridization function with substantial spectral weight near zero energy. In addition, as demonstrated in Sec.~\ref{subsec:noninteracting} and Sec.~\ref{subsec:Kanamori}, the required TT rank for high expansion orders can be significantly larger than that for low orders, posing a major computational challenge for TT-based impurity solvers.

\section{Conclusions}\label{sec:conclusion}
The integration and summation scheme based on tensor-train representations of Feynman diagrams is a promising approach for efficient and accurate quantum impurity solvers. In this work, we present an algorithm combining imaginary-time inchworm expansion with tensor-train methods that handles general interactions and off-diagonal hybridizations, with ergodicity problems mitigated through an adaptive random noise technique. Our results demonstrate that for moderate expansion orders, the tensor-train ranks remain computationally tractable, establishing this approach as a viable alternative to conventional quantum impurity solvers.

We identify three principal challenges requiring further methodological advancement. First, converged Green's functions for certain impurity problems necessitate exceptionally fine discretization of the bold propagator. Recent developments employing sum-of-exponentials representations of the bold propagator \cite{Jason_Exponentials_2024,Huang_Exponentials_2025} suggest promising solutions. Systematic integration of these approaches \cite{Mejuto_PES_2020,Jason_DLR_2022,Huang_PES_2023,Huang_Exponentials_2025,Zhang_Minipole_2024_a,Zhang_Minipole_2024_b,Zhang_Minipole_2025} with the inchworm algorithm warrants investigation. Alternatively, integrating tensor-train techniques into the self-consistent bold-line framework—where such representations for propagators appear more natural~\cite{Jason_Exponentials_2024,Huang_Exponentials_2025}—may be worth exploring.
Second, multi-orbital systems exhibit tensor-train ranks that grow substantially more rapidly at high expansion order compared to single-orbital models,
because of the enlarged configuration space from the spin-orbital indices in tensor trains. Further algorithmic advances in rank reduction could improve efficiency, as the current tensor-cross interpolation algorithm does not guarantee theoretically optimal ranks for a given precision.
Third, the bold expansion exhibits slow convergence with respect to the maximum expansion order when the impurities are coupled to baths with significant spectral weight near zero frequency. A deeper understanding of this issue and potential remedies remains an important direction for future research.

\section{Acknowledgment}
We thank  Sergei Iskakov, Wei-Ting Lin, Yuanran Zhu, Alec Dektor, Chao Yang, Jason Kaye, Miles Stoudenmire, Hirone Ishida, and Hiroshi Shinaoka for helpful discussions. 
Y.Y. and E.G. were supported by the National Science Foundation under Grant No. NSF DMR 2401159. Y.Y. acknowledges support from a predoctoral fellowship at the Flatiron Institute. 
A.E.\ was supported by the U.S.\ Department of Energy, Office of Science, Office
of Advanced Scientific Computing Research and Office of Basic Energy Sciences, Scientific Discovery through Advanced Computing (SciDAC) program under Award No.
DE-SC0022088.
D.Z. was supported by the National Science Foundation under Grant No. CHE-2154152.
G.C.\ acknowledges support by the ISF (Grant No. 2902/21), by the PAZY foundation (Grant No. 318/78) and by MOST NSF-BSF (Grant No. 2023720).
This work used Stampede3 at TACC through allocation PHY240286 from the Advanced Cyberinfrastructure Coordination Ecosystem: Services \& Support (ACCESS) program~\cite{access}, which is supported by National Science Foundation Grants No. 2138259, No. 2138286, No. 2138307,  No. 2137603, and No. 2138296. The Flatiron Institute is a division of the Simons Foundation.

The code used in this work is built on top of the TRIQS library~\cite{Parcollet_Triqs_2015,Charlebois_Triqs} and the xfac library~\cite{Nunez_xfac_2025}.

\section{DATA AVAILABILITY}
The data that support the findings of this article are openly available~\cite{yu_data_2025}.

\appendix
\section{Strong-coupling expansion and diagrammatic resummation frameworks}\label{appendix:inchworm_derivation}

\subsection{Strong-coupling expansion}\label{subsec:strong_coupling_expansion}
The partition function $Z_{\mathrm{imp}}$ of an impurity action $\mathcal{S}_{\mathrm{imp}} = \mathcal{S}_{\mathrm{loc}} + \sum_{1'1} \cc_{1'}\Delta_{1'1} c_1$ is given by the functional integral
\begin{equation}
    \begin{aligned}
        Z_{\mathrm{imp}}
        &=\int \mathcal{D}[\cc,c] \mathrm{e}^{-\mathcal{S}_{\mathrm{imp}}}\\
        &=\int \mathcal{D}[\cc,c] \mathrm{e}^{-\mathcal{S}_{\mathrm{\mathrm{loc}}}} \mathrm{e}^{-\sum\limits_{1'1} \cc_{1'}\Delta_{1'1} c_{1}}.
    \end{aligned}
\end{equation}
In the strong-coupling expansion framework, we expand the term $\mathrm{e}^{-\sum_{1'1} \cc_{1'}\Delta_{1'1} c_{1}}$ with power series, which leads to
\begin{equation}\label{eqn:bare}
\begin{aligned}
    Z_{\mathrm{imp}}
    &=\sum\limits_{n=0}^{\infty} \sum\limits_{\substack{1 \cdots n \\ 1'\cdots n'}} \int\mathcal{D}[\cc, c]\frac{\mathrm{e}^{- \mathcal{S}_{\mathrm{loc }}} }{(n!)^{2}}  c_{n}c ^{*}_{n'}  \cdots
    c_{1}\cc_{1'} \operatorname{det} \mathbf{\Delta}\\
    &=\sum\limits_{n=0}^{\infty} \sum\limits_{\substack{1 \cdots n \\ 1'\cdots n'}} 
    \frac{\operatorname{det} \mathbf{\Delta}}{(n!)^{2}}\operatorname{Tr} \left[\mathrm{e}^{-\beta \hat{H}_{\mathrm{loc}}}\mathcal{T} \hc_{n} \hcd_{n'}  \cdots
    \hc_{1}\hcd_{1'}
    \right].
\end{aligned}
\end{equation}
Here, $n$ is the expansion order, which corresponds to the number of hybridization functions. The determinant $\operatorname{det} \mathbf{\Delta} \equiv \sum\limits_{P\in S_{n}} (-1)^{n_{P}} \Delta_{1'P(1)} \cdots \Delta_{n'P(n)}$, where $P$ runs over all permutations $S_{n}$ of the set $\left\{1,\cdots,n\right\}$ and $n_{P}$ counts the number of exchanges in $P$. The second line of Eq.~(\ref{eqn:bare}) is expressed in second quantization form, where $\mathcal{T}$ is the ordering operator for imaginary time and $\hc^{(\dagger)}_{i} \equiv \mathrm{e}^{ \hat{H}_{\mathrm{loc}}\tau}\hc^{(\dagger)}_{\iota_{i}}\mathrm{e}^{- \hat{H}_{\mathrm{loc}}\tau}$. In practice, operators are represented as (block diagonalized) matrices in a local many-body basis, and the trace operation is carried out in this basis. Monte Carlo sampling of the high-dimensional integrals and sums in Eq.~(\ref{eqn:bare}) leads to the continuous-time hybridization expansion (CT-HYB) method~\cite{Werner_CTHYB_2006,Werner_CTHYB2_2006,Haule_CTHYB_2007,Gull_CTQMC_2011}.

Equation~(\ref{eqn:bare}) can be reformulated by explicitly enforcing time-ordering, yielding
\begin{equation}\label{eqn:bare2}
\begin{aligned}
Z_{\mathrm{imp}}
=&\sum\limits_{n=0}^{\infty} \sum_{\Phi}\sum\limits_{1\cdots2n}       
\sum_{\rm{D}} (-1)^{\sigma} (\Delta)^{n}
\operatorname{Tr} \left[\mathrm{e}^{-\beta \hat{H}_{\mathrm{loc}}} \hc_{2n}^{\phi_{2n}}  \cdots
    \hc_{1}^{\phi_{1}}
    \right]\\
=&\sum\limits_{n=0}^{\infty} \sum_{\Phi}\sum\limits_{1\cdots2n} 
\sum_{\rm{D}} (-1)^{\sigma} (\Delta)^{n} \\
&\operatorname{Tr} \left[\hat{R}^{0}(\beta-\tau_{2n}) \hc_{\iota_{2n}}^{\phi_{2n}} \cdots \hat{R}^{0}(\tau_{2}-\tau_{1})
    \hc_{\iota_{1}}^{\phi_{1}} \hat{R}^{0}(\tau_{1})
    \right],
\end{aligned}
\end{equation}
where the time variables are ordered as $0<\tau_1<\tau_2<\cdots < \tau_{2n}<\beta$, and the time integration takes the form    $\int_{0}^{\beta}\mathrm{d}\tau_{1}\int_{\tau_1}^{\beta}\mathrm{d}\tau_{2}\cdots \int_{\tau_{2n-1}}^{\beta}\mathrm{d}\tau_{2n}$.
The operator type is specified by $\phi_{i}\in \left\{-,+\right\}$, with $\hc^{-}\equiv \hc$ and $\hc^{+}\equiv\hcd$. The set $\Phi=\left\{\phi_1,\cdots,\phi_{2n}\right\}$ runs over all $\binom{2n}{n}$ possible choices of $n$ creation and $n$ annihilation operators. 
The notation $(\Delta)^{n}$ refers to a product of $n$ hybridization functions. The index $\mathrm{D}$ denotes all distinct assignments of the indices $\left\{1,2,\cdots,2n\right\}$ to the $n$ hybridization functions $\Delta$, when $\Phi$ is fixed. In the second line of Eq.~(\ref{eqn:bare2}), we introduce the bare  propagator $\hat{R}^{0}(\tau)\equiv \mathrm{e}^{-\hat{H}_{\mathrm{loc}}\tau}$.

The prefactor $(-1)^{\sigma}$ in Eq.~(\ref{eqn:bare2}) is determined as follows:  First, determine the number of exchanges $n_{c}$ required to rearrange the operator chain $\hc_{2n}^{\phi_{2n}} \hc_{2n-1}^{\phi_{2n-1}}  \cdots \hc_{1}^{\phi_{1}}$ into the form $\hc \hcd \hc \hcd \cdots \hc \hcd$. Second, define a mapping from $\left\{1,\cdots,2n\right\}$ to $\left\{1,1',\cdots,n,n'\right\}$ such that the operator chain $\hc \hcd \hc \hcd \cdots \hc \hcd$ with indices belong to $\left\{1,...,2n\right\}$ is mapped as  $\hc_{n} \hcd_{n'}  \cdots
\hc_{1}\hcd_{1'}$. Under this mapping, $(\Delta)^{n}$ is transformed into the form $\Delta_{1'P(1)} \cdots \Delta_{n'P(n)}$, and the number of exchanges $n_{P}$ for the permutation $P$ of the set $\left\{1,\cdots,n\right\}$ is computed. The exponent is then given by $\sigma = n_{c} + n_{P}$. Although the parity of $(-1)^{n_{c}}$ is not uniquely defined, it is easy to verify that the parity of $(-1)^{\sigma}$ is unambiguous.

\begin{figure}[htb]
\centering
\includegraphics[width=1\linewidth]{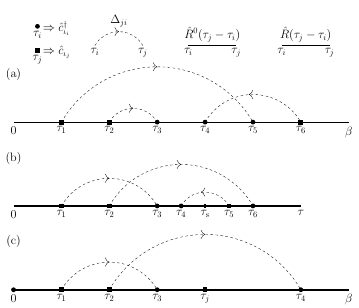}
\caption{Schematic of the contributions to the hybridization expansion, depicted as time-ordered diagrams. (a) Third-order bare expansion diagram for the partition function $Z_{\mathrm{imp}}$. (b) Third-order inchworm expansion diagram for the bold propagator $\hat{R}(\tau)$ with $\tau_{\mathrm{s}}$ denoting the ``split" time. (c) Third-order bold expansion diagram for the one-body Green’s function $G_{j0}$.}
\label{fig:hyb}
\end{figure}

The chain of operators in the trace of Eq.~(\ref{eqn:bare2}) and $(\Delta)^{n}$, along with the prefactor $(-1)^{\sigma}$, admits a diagrammatic representation. For example, in Fig.~\ref{fig:hyb}(a), we exemplify the diagramatic formulation for an order $n=3$ contribution to the expansion with, $\phi_{1},\phi_{2},\phi_{6}=-$, $\phi_{3},\phi_{4},\phi_{5}=+$ and $(\Delta)^{3} = \Delta_{32}\Delta_{46}\Delta_{51}$. 
The prefactor $(-1)^{\sigma}$ is determined as follows. First, $\hc_{6}\hcd_{5}\hcd_{4}\hcd_{3}\hc_{2}\hc_{1}$ can be reordered in the format of $\hc_{6}\hcd_{5}\hc_{2}\hcd_{4}\hc_{1}\hcd_{3}$ with $n_{c}=3$. The mapping $(6\to 3,\;5\to 3',\;2\to 2,\;4\to 2',\;1\to 1,\;3\to1')$ restores the form $\hc_{3}\hcd_{3'}\hc_{2}\hcd_{2'}\hc_{1}\hcd_{1'}$ while transforming $\Delta_{32}\Delta_{46}\Delta_{51}$ to $\Delta_{1'2}\Delta_{2'3}\Delta_{3'1}$, yielding $n_{P}=2$. Therefore, $(-1)^{\sigma}=(-1)^{3+2}=-1$. In the diagrammatic language, the prefactor $(-1)^{\sigma}$ can alternatively be determined by counting the number of crossings for the hybridization lines $n_{\mathrm{crossing}}$ and the number of forward-oriented hybridization lines $n_{\mathrm{forward}}$ (those aligned with the contour from $0$ to $\beta$). The relation $(-1)^{\sigma}=(-1)^{n_{\mathrm{crossing}}+n_{\mathrm{forward}}}$ then follows. In Fig.~\ref{fig:hyb}(a), $n_{\mathrm{crossing}}=1$ and $n_{\mathrm{forward}}=2$, which gives $(-1)^{\sigma}=(-1)^{1+2}=-1$.

\subsection{Diagrammatic resummation frameworks}\label{subsec:resummation}
The bare hybridization expansion is defined by Eq.~(\ref{eqn:bare}) or Eq.~(\ref{eqn:bare2}), with the bare propagator $\hat{R}^{0}$ serving as the fundamental building block.
In physically interesting regimes, this approach often requires large expansion orders in order to reach convergence. 
Various approaches for regrouping diagrams exist that enable a significant reduction of the expansion order~\cite{Prokof'ev_Bold_2008,Haule_bold_2001,Gull_Bold_2010,gull_numerically_2011,Ruegg_XCA_2013,cohen_numerically_2013,cohen_greens_2014,Guy_Bold_2014,Guy_Inchworm_2015,Eitan_inchworm_2020,Strand_Inchworm_2024,Jason_Exponentials_2024,Huang_Exponentials_2025}. These approaches leverage a bold propagator $\hat{R}(\tau)$, defined as
\begin{equation}\label{eqn:propagator}
\begin{aligned}
\hat{R}(\tau)
&=\sum\limits_{n=0}^{\infty} \sum_{\Phi}\sum\limits_{1\cdots2n} 
\sum_{\mathrm{D}} (-1)^{\sigma} (\Delta)^{n} \\
&\left[\hat{R}^{0}(\tau-\tau_{2n}) \hc_{\iota_{2n}}^{\phi_{2n}} \cdots  \hat{R}^{0}(\tau_{2}-\tau_{1})
    \hc_{\iota_{1}}^{\phi_{1}}\hat{R}^{0}(\tau_{1})
    \right],
\end{aligned}
\end{equation}
which is subject to the time-ordering $0<\tau_{1}<\tau_{2}<\cdots\tau_{2n}<\tau$, and the time integration takes the form    $\int_{0}^{\tau}\mathrm{d}\tau_{1}\int_{\tau_1}^{\tau}\mathrm{d}\tau_{2}\cdots \int_{\tau_{2n-1}}^{\tau}\mathrm{d}\tau_{2n}$. By definition, $\hat{R}(0)$ is an identity matrix.
This expression is closely related to Eq.~(\ref{eqn:bare2}) and one can easily verify $Z_{\mathrm{imp}}=\operatorname{Tr}[\hat{R}(\beta)]$. 

In the inchworm algorithm, the value of $\hat{R}$ at $\tau$ is computed under the assumption that $\hat{R}$ is already known for all times between $0$ and $\operatorname{max}(\tau_{\mathrm{s}},\tau -\tau_{\mathrm{s}})$, where $\tau_{\mathrm{s}}<\tau$ denotes a ``split" time. The bold propagator at $\tau$ is given by the expression
\begin{widetext}
    \begin{equation}\label{eqn:propagator2}
    \begin{aligned}
    \hat{R}(\tau)
    &=\sum\limits_{n=0}^{\infty} \sum_{\Phi}\sum\limits_{1\cdots2n} \sum_{\mathrm{D}_{\mathrm{p}}} (-1)^{\sigma} (\Delta)^{n} 
    \left[\hat{R}(\tau-\tau_{2n}) \hc_{\iota_{2n}}^{\phi_{2n}}\cdots \hat{R}(\tau_{q+1}-\tau_{\mathrm{s}}) \hat{R}(\tau_{\mathrm{s}}-\tau_{q}) \cdots \hat{R}(\tau_{2}-\tau_{1})
    \hc_{\iota_{1}}^{\phi_{1}}\hat{R}(\tau_{1})
    \right]\\
    &\equiv\sum\limits_{n=0}^{\infty} \sum_{\Phi}\sum\limits_{1\cdots2n} \sum_{\mathrm{D}_{\mathrm{p}}} \underline{\hat{R}}(\tau|n;\Phi;\iota_{1}\cdots\iota_{2n};\tau_{1}\cdots\tau_{2n};\mathrm{D}_{\mathrm{P}}),
    \end{aligned}
\end{equation}
\end{widetext}
where $0<\tau_{1}<\cdots <\tau_{2n}<\tau $ and $\tau_{q}\leq \tau_{\mathrm{s}}\leq \tau_{q+1}$ for certain $q$. Here, $\mathrm{D}_{\mathrm{p}}$ denotes the set of all diagrams in which every connected cluster of hybridization lines crosses the split time $\tau_{\mathrm{s}}$, a property referred to as ``inchworm proper". The number of proper diagrams exhibits factorial scaling with expansion order. This scaling can be reduced to exponential through summation schemes based on the inclusion--exclusion principle~\cite{Boag_IE_2018} or similar algorithms \cite{rossi_determinant_2017}. Fig.~\ref{fig:hyb}(b) shows a third-order diagram where $\Delta_{31}$ and $\Delta_{62}$ form one connected cluster and $\Delta_{45}$ forms another, both crossing $\tau_{\mathrm{s}}$. The sign of the diagram is $(-1)^{\sigma}=(-1)^{n_{\mathrm{crossing}}+n_{\mathrm{forward}}}=(-1)^{1+2}=-1$. One can check the validity of Eq.~(\ref{eqn:propagator2}) by substituting the bold propagator $\hat{R}$ on the right hand side of Eq.~(\ref{eqn:propagator2}) with its bare expansion Eq.~(\ref{eqn:propagator}) in terms of $\hat{R}^{0}$ and hybridization lines $\Delta$.  This substitution recovers all valid diagram topologies in the bare expansion of Eq.~(\ref{eqn:propagator}) without generating any spurious terms not included in it.

After obtaining the approximated bold propagator according to the inchworm process introduced in Sec.~\ref{subsec:inchworm}, the one-body Green's function,
\begin{equation}
\begin{aligned}
    G_{j0} &\equiv - \braket{c_{j} \cc_{0}}\\
    &= -\frac{1}{Z_{\mathrm{imp}}} \int \mathcal{D}[\cc,c] \mathrm{e}^{-\mathcal{S}_{\mathrm{imp}}} c_{j} \cc_{0} ,
\end{aligned}
\end{equation}
can be evaluated using the bold propagator \cite{Guy_Bold_2014} via
\begin{widetext}
\begin{equation}\label{eqn:gf}
\begin{aligned}
G_{j0}
& =-\frac{1}{Z_{\mathrm{imp}}} \sum\limits_{n=0}^{\infty} \sum_{\Phi}\sum\limits_{1\cdots2n} 
\sum_{\mathrm{D}_{\mathrm{p}}} (-1)^{\sigma+q} (\Delta)^{n} 
\operatorname{Tr}\left[\hat{R}(\beta-\tau_{2n}) \hc_{\iota_{2n}}^{\phi_{2n}} \cdots \hat{R}(\tau_{q+1}-\tau_{j})\hc_{\iota_{j}} \hat{R}(\tau_{j}-\tau_{q}) \cdots \hat{R}(\tau_{2}-\tau_{1})
    \hc_{\iota_{1}}^{\phi_{1}} \hat{R}(\tau_{1}) \hcd_{\iota_{0}} \right]\\
& \equiv -\frac{1}{Z_{\mathrm{imp}}} \sum\limits_{n=0}^{\infty} \sum_{\Phi}\sum\limits_{1\cdots2n} 
\sum_{\mathrm{D}_{\mathrm{p}}} \underline{G}_{j0}(n;\Phi;\iota_{1}\cdots\iota_{2n};\tau_{1}\cdots\tau_{2n};\mathrm{D}_{\mathrm{P}}).
\end{aligned}
\end{equation}
\end{widetext}
The derivation follows the derivation of the inchworm expansion of the bold propagator in Eq.~(\ref{eqn:propagator2}). The proper diagram is defined with respect to $\tau_{j}$ instead of $\tau_{\mathrm{s}}$ and $q$ counts the number of creation or annihilation operators between $\hc_{\iota_{j}}$ and $\hcd_{\iota_{0}}$. In Fig.~\ref{fig:hyb}(c), we illustrate a third-order diagram for the Green's function. Here, the number of crossing $n_{\mathrm{crossing}}=1$ and the number of forward-oriented hybridization lines is $n_{\mathrm{forward}}=2$, the number of creation or annihilation operators between $0$ and $\tau_{j}$ is $q=3$. Therefore, the prefactor is $(-1)^{1+2+3}=1$.

With fixed expansion order $n$, the computational cost of Eqs.~\eqref{eqn:propagator2} and \eqref{eqn:gf} mainly comes from the factorial number of diagram topologies $\mathrm{D}_{\mathrm{p}}$ and the high-dimensional integration over $2n$ variables $\left\{\tau_{1},\cdots,\tau_{2n}\right\}$. The summation over the discrete indices $\left\{\iota_{1},\cdots,\iota_{2n}\right\}$ scales as $N_{\iota}^{2n}$, where each index has dimension $N_{\iota}$. Finally, the cost of enumerating $\Phi$ scales as $\binom{2n}{n}<\sum_{x=0}^{2n} \binom{2n}{x}=2^{2n}\leq N_{\iota}^{2n}$, for $N_{\iota}>1$. 

\section{Ergodicity Problem}\label{appendix:ergodicity}
\begin{figure}[htb]
    \centering
    \includegraphics[width=0.95\linewidth]{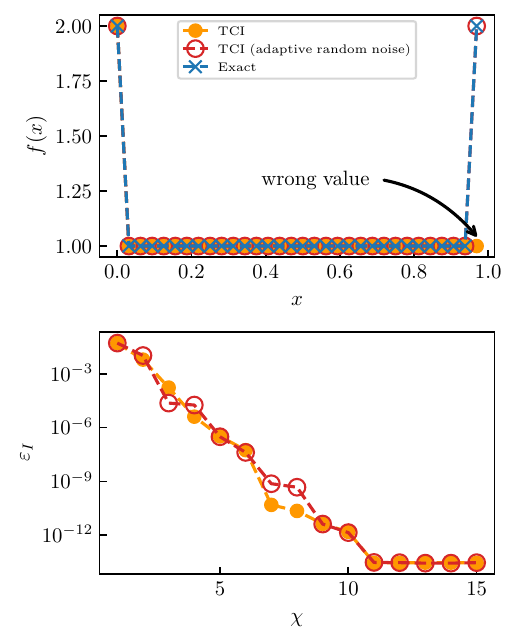}
    \caption{Illustration of the ergodicity problem and the proposed solution with adaptive random noise. Top panel: A representative function $f(x)$ where standard TCI (orange filled circles) fails to capture the correct TT representation, compared to the TCI with adaptive random noise (red open circles) and the exact values (blue crosses). Bottom panel: Absolute integration error for an integrand without ergodicity problems as a function of TT rank $\chi$, comparing the standard TCI (orange filled circles) and the proposed TCI algorithm with adaptive random noise (red open circles).}
    \label{fig:ergodicity}
\end{figure}
Despite the ability of TCI to accurately describe many functions upon increasing the rank of TT, the current version of the algorithm may fundamentally fail to capture certain functions. A representative example is illustrated in the top panel of Fig.~\ref{fig:ergodicity}, where we discretize $x$ in the interval $[0,1]$ using a quantics representation, $x=\sum_{i=1}^{N} \sigma_{i}/2^{i}$, with each physical index $\sigma_{i}$ in TT taking binary values $0$ or $1$. The target function is defined as $f(x)=a$ if all indices are $1$ or all are $0$, and $f(x)=b$ otherwise. Here, we take $a=2$, $b=1$, and $N=5$. The orange filled circles represent the function learned by TCI (initialized with all zeros as the starting pivot), while the blue crosses denote the exact values. Evidently, TCI fails to give the correct value of $f(x)$ when all indices are $1$. In fact, it can be shown that for $b\neq 0$ and $N \geq 5$, or for $b=0$ and $N \geq 3$, TCI consistently misses the value $a$ at the last discrete grid point.

The ergodicity problem is not limited to visually flat functions, as in the top panel of Fig.~\ref{fig:ergodicity}. Instead, the functions TCI failed to describe are generally characterized by the presence of low-rank domains in configuration space, which can also exhibit finite curvature visually. The presence of extended regions of constant values—especially zeros—in the target function is the most straightforward mechanism for the formation of these low-rank domains. TCI fails to accurately represent these functions because of its local update scheme for four-leg tensors [dashed box in Fig.~\ref{fig:tensor}(a)]. When exploring low-rank domains, the approximation criteria of four-leg tensors are easily satisfied, causing premature termination of pivot exploration. This results in incomplete sampling of configuration space and consequently yields an incorrect tensor train representation. In many cases, the low-rank domains can be exactly described by the TT decomposition. Consequently, regardless of how the approximation criteria of four-leg tensors are adjusted, the final results remain incorrect, which is the case in the top panel of Fig.~\ref{fig:ergodicity}. In other scenarios, adjusting the approximation criteria may mitigate this issue, but a delay in the exploration of the full configuration space could persist, leading to a significant slowdown in convergence.

References.~\cite{Nunez_xfac_2025,Ishida_Tensor_Phonon_2024} introduced the global pivot update method to address this issue. However, a valid global pivot is often not known {\it a priori} and must be determined algorithmically~\cite{Ishida_Tensor_Phonon_2024}, a process that can incur exponential computational costs. In this work, we propose an algorithm that employs adaptive random noise to mitigate the ergodicity problem, achieving this with almost no additional computational overhead compared to the original TCI algorithm.  The details of this algorithm are displayed in Algorithm~\ref{alg:tci_adaptive}. 
We note that the addition of the random noise~\cite{White_DMRG_1992,Chan_dmrg_2002,Zgid_dmrg_noise} during the algorithm execution is a standard numerical procedure in DMRG algorithms, where such noise is included to prevent the numerical procedure from getting stuck in a local optimization minimum. Such local minima are reached as a result of an earlier exclusion of blocks representing certain configurations that may have only a small weight during initial steps. The addition of a small amount of random noise ensures that there is always a path for reaching other minima numerically.

\begin{algorithm}[htb] 
 \caption{TCI with adaptive random noise}
 \label{alg:tci_adaptive}
 \textbf{Input:} Initial pivot error $\varepsilon^{\mathrm{init}}_{\mathrm{pivot}}\neq 0$, max sweeps $\mathit{max\_sweep}>1$, target function $f$, noise control parameter $\alpha$, random number generator $R_{[-1,1]}$\\
 \textbf{Output:} Optimized TT representation of target function $f$\;
 \vspace{0.1in}
 $\varepsilon_{\mathrm{pivot}} \leftarrow \varepsilon^{\mathrm{init}}_{\mathrm{pivot}}$\;
 $\mathit{has\_nonzero\_pivot} \leftarrow \mathrm{false}$\;
 Construct TT for the target function $f+g$ with an arbitrary initial pivot, where $g = \alpha \cdot \varepsilon_{\mathrm{pivot}} \cdot R_{[-1,1]}$\;
 \For{$i \leftarrow 1$ \KwTo $\mathit{max\_sweep}$}{
   \If{$i \neq \mathit{max\_sweep}$ \textbf{or} $\mathit{has\_nonzero\_pivot}$}{
     Optimize TT pivots for target function $f + g$ via TCI two-site updates~\cite{Nunez_xfac_2025}\;
   }
   \If{$i = \mathit{max\_sweep}$ \textbf{and} $!\mathit{has\_nonzero\_pivot}$}{
   \Return \text{``Warning: Target function is zero"}\;
   }
   Update $\varepsilon_{\mathrm{pivot}}$ according to the results of TCI two-site updates\;
   \If{$i = \mathit{max\_sweep}-1$}{
     $\varepsilon_{\mathrm{pivot}} \leftarrow 0$ \tcp*{Disable noise in the final sweep}
   }
   \If{$!\mathit{has\_nonzero\_pivot}$ \textbf{and} $\exists$ a pivot $p$ in TT with $f(p) \neq 0$}{
     $\mathit{has\_nonzero\_pivot} \leftarrow \mathrm{true}$\;
   }
 }
 \Return Optimized TT\;
\end{algorithm}

The key idea of the proposed algorithm is to modify the target function $f$ in each sweep of the TCI algorithm by adding an adaptive random noise $g$. Since the random noise $g$ is a maximal-rank function, the local updates for $f+g$ avoid premature termination. To ensure the TCI algorithm preferentially selects pivots based on $f$ rather than $g$, we constrain the noise magnitude to be small compared to those $f$ in regions not yet accurately described by TT. We find that the pivot error $\varepsilon_{\mathrm{pivot}}$, the maximum error of the local four-leg tensor from the previous sweep, provides a natural scale for this purpose, which decays with successive sweeps. Since $\varepsilon_{\mathrm{pivot}}$ has already been computed in the original TCI algorithm, the proposed algorithm introduces no additional cost. The noise is set as $g=\alpha\cdot\varepsilon_{\mathrm{pivot}}\cdot R_{[-1,1]}$, where $\alpha<1$ is a control parameter and $R_{[-1,1]}$ denotes a uniformly distributed random variable on the interval $[-1,1]$. Crucially, $R_{[-1,1]}$ generates distinct values for each unique combination of physical indices of TT. Empirically, the algorithm is robust to the choice of $\alpha$, and we fix $\alpha=0.1$ throughout this work. This adaptive random noise scheme enhances exploration of the configuration space by promoting pivot selection across a broader region, while its systematic decay with successive sweeps guarantees numerical accuracy. The incorporation of random noise also offers an additional advantage by guaranteeing that all potential pivots for the function $f+g$ remain non-zero. This enables straightforward initialization of the TT representation prior to TCI optimization, in contrast to the original TCI algorithm that may require additional steps to locate suitable non-zero initial pivots.

Using the proposed algorithm, we resolve the ergodicity problem discussed at the beginning of this section, as demonstrated by the open red circles in the top panel of Fig.~\ref{fig:ergodicity}. For further validation, we benchmark our method against a test case without ergodicity problems from Ref.~\cite{Nunez_xfac_2025}. Specifically, we evaluate the high-dimensional integral
\begin{equation}
    I = \int_{[0,1]^{5}} \mathrm{d}x_{1} \cdots \mathrm{d}x_{5} f(\bf{x}),
\end{equation}
where the integrand
\begin{equation}
    f(\mathbf{x}) = \frac{2^{5}}{1+2 \sum_{i=1}^{5} x_{i}}
\end{equation}
is represented as a TT with each variable $x_{i}$ treated as a physical index. The bottom panel of Fig.~\ref{fig:ergodicity} compares the absolute error $\varepsilon_{I}$ for the original TCI algorithm (orange filled circles) and the modified TCI algorithm with adaptive random noise (red open circles), plotted against the TT rank $\chi$. The adaptive random noise does not degrade pivot selection even at small
$\chi$, demonstrating the robustness of our approach.

While the proposed algorithm performs robustly in all cases studied in this work, it does not offer a rigorous theoretical guarantee for resolving ergodicity problems. A deeper understanding of ergodicity problems and potential solutions remains an important direction for future research.

\section{Propagator for the noninteracting limit}\label{appendix:propagator_noninteracting}
\begin{figure}[htb]
    \centering
    \includegraphics[width=1\linewidth]{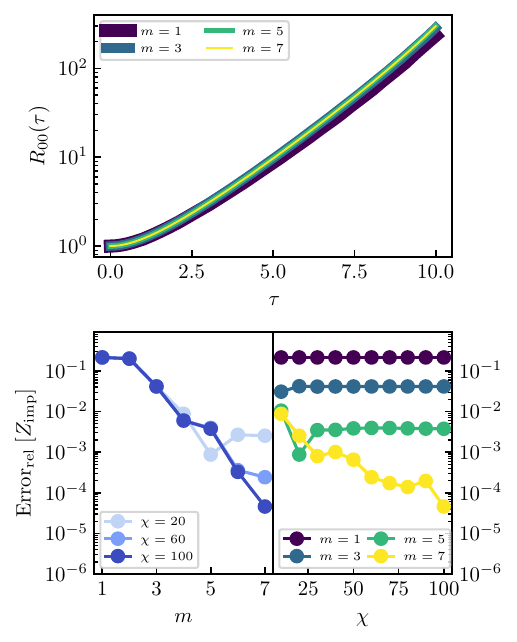}
    \caption{
    Bold propagators of the spinless non-interacting model. Top panel: First matrix element of the propagator $R_{00}(\tau)$ for different maximum expansion orders $m$ (color-coded). Bottom panel: Relative error of the partition function as a function of (left) maximum expansion order $m$ (color-coded by $\chi$) and (right) rank $\chi$ (color-coded by $m$).
    }
    \label{fig:spinless_Bethe_R}
\end{figure}

We present the bold propagator for the spinless non-interacting model discussed in Sec.~\ref{subsec:noninteracting}. The calculations are performed using parameters $N_{\mathrm{inch}}=11$. The top panel of Fig.~\ref{fig:spinless_Bethe_R} shows the first matrix element of the propagator $R_{00}(\tau)$ for different maximum expansion orders $m$. The bottom panel presents the relative error of the partition function, computed via $Z_{\mathrm{imp}}=\operatorname{Tr}[\hat{R}(\beta)]$, as a function of the maximum expansion order $m$ (left) and the TT rank $\chi$ (right). The exact partition function is given by
\begin{equation}
    Z^{\mathrm{exact}}=\exp\left[ \int_0^1 dx  \ln (1+e^{- 2 \beta \cos(2\pi x)}) \right].
\end{equation}
Similar to the case of the Green’s function discussed in Sec.~\ref{subsec:noninteracting}, a modest rank of $\chi \sim 20$ is sufficient for maximum expansion orders $m\leq 5$.

\section{Accurate description of  propagator}\label{appendix:linear-Cheybshev}
\begin{figure}[htb]
    \centering
    \includegraphics[width=0.95\linewidth]{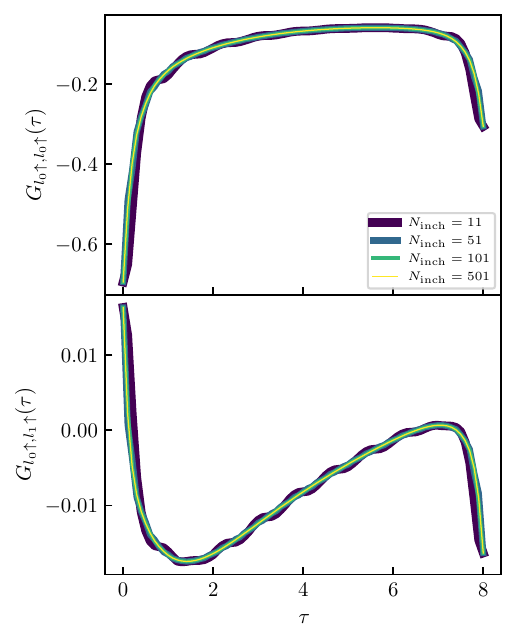}
    \caption{Green's functions of the two-orbital spinful model computed from the first-order expansion ($m=1$) using the exact bold propagator. Top panel: Diagonal component. Bottom panel: Off-diagonal component. Different colors correspond to different discretization grids for the exact bold propagator.}
    \label{fig:linear-Cheybshev}
\end{figure}
For the two-band model introduced in Sec.~\ref{subsec:Kanamori}, we find that the corresponding propagator requires a fine grid for an accurate description, even when using a linear-Chebyshev grid. In Fig.~\ref{fig:linear-Cheybshev}, the top and bottom panels display the diagonal and off-diagonal Green's functions obtained from a first-order expansion $m=1$, corresponding to zero hybridization lines, evaluated with the exact bold propagator. The different colors indicate the grid resolution used for the exact propagator. Unlike the spinless case discussed in Sec.~\ref{subsec:noninteracting} and Appendix~\ref{appendix:propagator_noninteracting}, where $N_{\mathrm{inch}}= 11$ are sufficient to describe the propagator accurately, here the same parameters still result in oscillatory artifacts. In fact, for this case, a much finer linear grid with $N_{\mathrm{inch}}\sim\mathcal{O}(10^{2})$ is required to achieve a converged and smooth Green's function. This imposes a significant computational cost on the inchworm calculation of the propagator, as it implies that the number of inchworm steps needs to exceed $\mathcal{O}(10^{2})$.

\bibliography{references}
\end{document}